\documentclass[useAMS,usenatbib]{mn2e}

\usepackage{color}

\usepackage{graphicx}

\newcommand{\gps}{\ensuremath{g_{\rm P1}}}
\newcommand{\rps}{\ensuremath{r_{\rm P1}}}
\newcommand{\ips}{\ensuremath{i_{\rm P1}}}
\newcommand{\zps}{\ensuremath{z_{\rm P1}}}
\newcommand{\yps}{\ensuremath{y_{\rm P1}}}

\newcommand{\grizy}{\gps, \rps, \ips, \zps, \yps}
\newcommand{\PS}{\protect\hbox{Pan-STARRS1} }

\title[The \PS Small Area Survey 2]{The  \PS Small Area Survey 2}
\author[N. Metcalfe et al.]{\parbox{\textwidth}{N. Metcalfe$^{1}$\thanks{E-mail:nigel.metcalfe@durham.ac.uk}, D.~J. Farrow$^{2}$, S. Cole$^{2}$, P.~W. Draper$^{1,2}$, P. Norberg$^{2}$,  
W.~S. Burgett$^{3}$,
K.~C. Chambers$^{3}$,
L. Denneau$^{3}$,
H. Flewelling$^{3}$,
N. Kaiser$^{3}$,
R. Kudritzki$^{3}$,
E.~A. Magnier$^{3}$,
J.~S. Morgan$^{3}$,
P.~A. Price$^{4}$,
W. Sweeney$^{3}$,
J.~L. Tonry$^{3}$,
R.~J. Wainscoat$^{3}$,
C. Waters$^{3}$
}
\vspace{0.4cm}\\
\parbox{\textwidth}
{$^{1}$Dept. of Physics, Univ. of Durham, South Road, Durham DH1 3LE, UK\\
$^{2}$Institute for Computational Cosmology, Dept. of Physics, Univ. of Durham, South Road, Durham DH1 3LE, UK\\
$^{3}$Institute for Astronomy, University of Hawaii, 2680 Woodlawn Drive, Honolulu HI 96822\\
$^{4}$Department of Astrophysical Sciences, Princeton University, Princeton, NJ 08544, USA\\
}}

\begin{document}

\date{}

\pagerange{\pageref{firstpage}--\pageref{lastpage}} \pubyear{2013}

\maketitle

\label{firstpage}

\begin{abstract}
The \PS survey is acquiring  multi-epoch imaging in 5 bands
(\grizy) over the entire sky north of declination $-30\deg$ (the $3\pi$
survey). In July 2011 a test area of about 70 sq.deg. was observed to the 
expected final depth of the main survey. In this, the first of a series of 
papers 
targetting the galaxy count and clustering properties of the combined
multi-epoch test area data, we present a 
detailed investigation into the depth 
of the survey and the reliability of the \PS analysis software. 
We show that the \PS reduction software can recover the properties of 
fake sources, and show good agreement between the magnitudes 
measured by \PS and those from Sloan Digital Sky Survey. 
We also examine the number of false detections apparent in the \PS data. 
Our comparisons show that
the test area survey is somewhat deeper than the
Sloan Digital Sky Survey in all bands, and, in particular, the $z$ band
approaches the depth of the stacked Sloan Stripe 82 data.

\end{abstract}

\begin{keywords}
surveys -- catalogues -- methods: data analysis -- techniques: image processing.
\end{keywords}

\section{Introduction}
\label{sec:intro}

The \PS (hereafter PS1) system \citep{PS1_system} is a 1.8-m aperture, f/4.4 telescope (Hodapp et al. 2004) 
illuminating a 1.4 Gpixel detector spanning a $3.3\deg$ field of 
view (\citealt{onaka} and \citealt{PS1_GPCA}), sited at the Haleakala 
Observatory on the island of Maui in Hawaii, and 
dedicated to sky survey observations. 
PS1 is undertaking a number of surveys, but the largest is the $3\pi$ survey (Chambers et al., in preparation), 
which is scanning the entire sky north of declination $-30\deg$ in five
filters, \grizy \ \citep{photo}, in 6 separate epochs spanning $\sim3.5$ years, 
each epoch consisting of a pair of exposures taken $\sim25$ minutes apart. 
By stacking
all these exposures, PS1 will provide a 30,000 sq.deg. survey of the sky to
a depth expected to be somewhat greater than that of the Sloan Digital Sky Survey \citep[SDSS;][]{sdss}, especially at redder wavelengths. The survey is expected to be completed early in 2014 and publicly 
released to the world by the end of that year.

In order to demonstrate the capabilities of the $3\pi$ survey, 
and to act as a test area to highlight any potential issues with the data
reduction and analysis, a demonstration area with the full number of exposures (12) at 
each pointing in each band (\gps, \rps, \ips, \zps, \yps) was undertaken with 
the PS1 telescope. This is known as the Small Area Survey version 2, 
hereafter SAS2 (version 1 was a similar test survey taken on a different area of sky a year previous to this).

This is the first of a series of papers whose aim is to demonstrate the 
viability of galaxy clustering studies on the stacked $3\pi$ survey by testing
the properties of SAS2.
In this paper we concentrate on more general issues of the data 
and subject the PS1 reduction software to a rigorous 
investigation, with emphasis on the depth of the stacked survey. We test the data analysis 
software ({\sc psphot}) on fake sources, as well as on SDSS fields, and then compare the
SAS2 source counts with the SDSS DR8 \citep{sdss8} and Stripe 82 \citep{s82} catalogues in this region. For the $y$-band, where there is no SDSS data,
we compare with the UKIDSS LAS \citep{UKIDSS}, and with deeper PS1 data.

In \citet[hereafter Paper II]{paper2} we will turn our attentions more specifically to galaxies, and investigate the 
counts and clustering on the SAS2, paying particular regard to the 
variable depth of coverage on small scales, which is an unavoidable 
feature of the PS1 camera and observing strategy.

\begin{figure}
\begin{center}
\centerline{\includegraphics[width=3.75in]{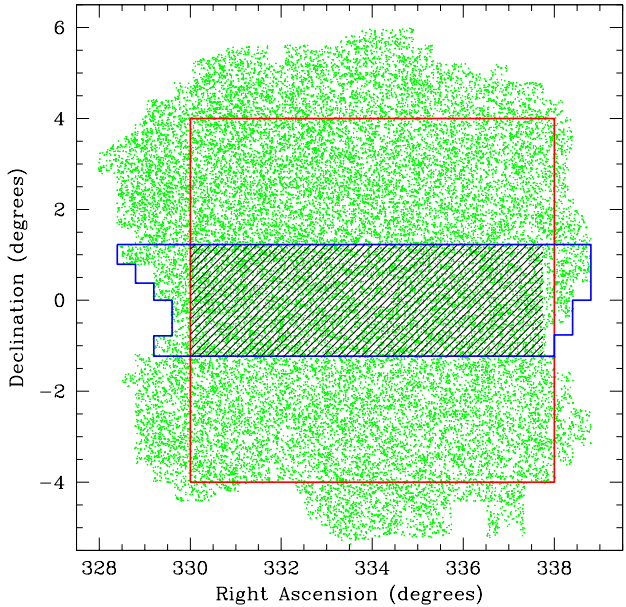}}
\caption{The angular distribution of objects in the SAS2 \rps-band source catalogues (green points).
The extent of our overlapping Stripe 82 (blue line) and DR8 (red line) catalogues is also shown.
The hatched zone represents the area in common to all three surveys. Coordinates are J2000.}
\label{fig:coverage}
\end{center}
\end{figure}

\section{The Small Area Survey}
\label{sec:sas2}

\begin{figure}
\begin{center}
\centerline{\includegraphics[width=3.5in]{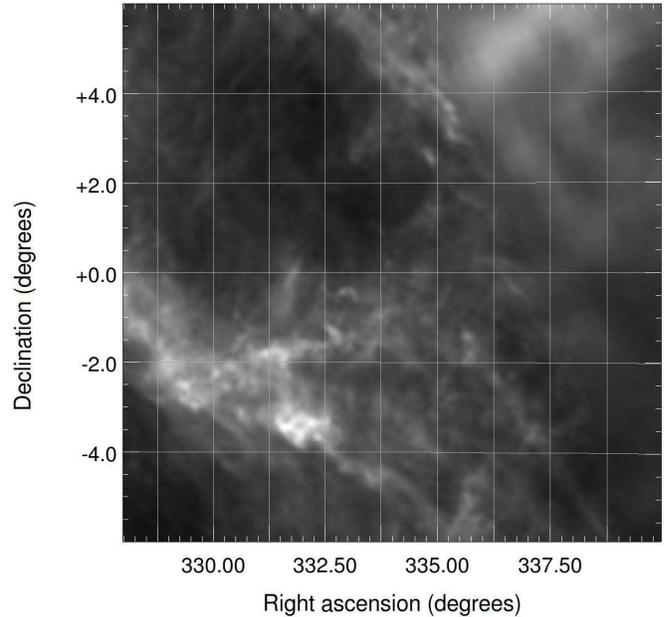}}
\caption{Greyscale map of E(B-V) reddening in the SAS2 area from the NASA/IPAC Infrared Science Archive (http://irsa.ipac.caltech.edu/applications/DUST/), derived from \citet{s98}. Black corresponds to $E(B-V)=0.01$, white to $E(B-V)=0.25$~mag.}
\label{fig:dust}
\end{center}
\end{figure}

SAS2 was observed over 2 nights in
the week after new moon at the beginning of July 2011. As with the
real $3\pi$ survey, the exposures were split into 6 pairs of observations at
different rotation angles on the sky. Ideally, each patch of sky  sees 
a total of 12 exposures, which are then stacked to produce a deeper
image. The survey is centred on
$22^{\rm h}15^{\rm m}$ RA, $0{\degr}00'$ Dec. (J2000), and the area of full 
coverage encompasses roughly
an $8\degr\times8\degr$ square, with a further $~1\degr$ wide strip around the
edge with reduced coverage due to dithering. There are a total of 124 
individual exposures 
in each band. The SDSS DR8 catalogue covers this area, whilst
there is substantial overlap with the deeper SDSS Stripe 82 catalogue. 
Fig. \ref{fig:coverage} shows the distribution
of detected sources on the sky from the stacked SAS2, together with the
areas we used from the SDSS DR8 and Stripe 82 
surveys (mainly restricted to areas with full SAS2 coverage in all 5 bands) . 
Our intercomparisons with SDSS are mainly based on 
the area in common to 
all three surveys ($\sim16$ sq.deg.), shown as the hatched zone on 
Fig. \ref{fig:coverage}.
Fig. \ref{fig:dust} shows the reddening distribution across the field. The
mean $E(B-V)$ is $\sim0.08$ mag, implying an \rps\ extinction of 
$\sim0.2$~mag, although $E(B-V)$ rises much higher to $\sim0.26$ mag around 
$22^{\rm h} 08^{\rm m}$ RA, $-3{\degr} 30'$ Dec.

A brief description of the PS1 camera is appropriate here. Full details can
be found in \citet{onaka} and \citet{PS1_GPCA}.
The detector consists
of 60 Orthogonal Transfer Arrays (OTA), about 4800 pixels square, arranged 
in an $8\times8$ pattern,  excluding the four corners. OTAs are reduced 
independently by the PS1 pipeline software, but a few operations, e.g. 
photometric zero-pointing, are performed globally across the exposure.
Each OTA itself consists of $8\times8$ CCD cells about 600 pixels 
across, with an image scale of $\sim0.26$~arcsec~pixel$^{-1}$ (the exact scale 
varies with position on the camera by about 1 per cent). There 
are gaps between each cell of between $6$ and $8$~arcsec and larger gaps 
between the OTAs of about $36$~arcsec in one direction and $70$~arcsec
in the other. As a result of this, and due to the dithering employed, when 
PS1 exposures are stacked, the resulting 
coverage can be very inhomogeneous.

Given the SAS2 is meant to be a demonstration of the $3\pi$ survey as a 
whole, the obvious question which arises is how do the properties, 
in particular seeing and sky brightness, of SAS2 compare with 
the wider $3\pi$ survey. Table \ref{table:drm} shows the statistics 
of the individual exposures which compose the SAS2. The zeropoints 
give the magnitude on the PS1 native system  of one count (ADU) per second
on the detector \citep[for a description of how the PS1 data are photometrically calibrated, see][]{ubercal}. The sky brightness and FWHM of the 
point spread function (PSF) 
come from the average value of model fits to the whole area of each individual 
exposure provided by the PS1 pipeline processing (the FWHM, in
particular, varies with position in the focal plane). SAS2 is, in fact, 
extremely uniform in its properties. The rms scatter in the sky value 
is $\le0.1$~mag in all bands,  
whilst for FWHM it is $\le0.05$~arcsec. The zero points scatter by 
$<0.01$~mag within each band.

\begin{table}
\setlength{\tabcolsep}{3pt}
\caption{Statistics of the SAS2 individual, i.e. unstacked, exposures. For comparison, the
values in parenthesis for the sky brightness and FWHM represent those for all the $3\pi$ exposures taken by October 2012.}
\begin{center}
\begin{tabular}{cllllll}
\hline
\hline
{\bf Filter} & {\bf Median}&{\bf Median sky}&{\bf Median} & {\bf Exp.} & {\bf Mean}\\
&{\bf zero pt.}&{\bf brightness}&{\bf FWHM}&{\bf time}&{\bf Airmass}\\
&(mag)&(mag/$''^{2}$)&($''$)&(s)&\\
\hline
\gps  &  $24.44$ & $21.96$ $(21.95)$ & $1.06$ $(1.33)$ & $43$ &1.25\\
\rps  &  $24.66$ & $21.12$ $(20.88)$ & $0.94$ $(1.19)$ & $40$ &1.15\\
\ips  &  $24.57$ & $20.45$ $(19.79)$ & $0.93$ $(1.13)$ & $45$ &1.09\\
\zps  &  $24.23$ & $19.65$ $(19.17)$ & $0.87$ $(1.08)$ & $30$ &1.07\\
\yps  &  $23.25$ & $18.46$ $(18.30)$ & $0.86$ $(0.99)$ & $30$ &1.08\\
\hline
\end{tabular}
\end{center}
\label{table:drm}
\end{table}

The values for sky brightness and seeing are somewhat better than for
the current $3\pi$ survey data (to October 2012), which are listed
in parenthesis in Table \ref{table:drm}.
The $3\pi$ data, of course, are a much more
heterogeneous sample, and the FWHM figures are skewed somewhat by a 
long tail to high values. The brighter skies in the redder
bands are due to the fact that these bands are usually scheduled
nearer full moon than was the case for SAS2.
Fig. {\ref{fig:rseeing} shows the histogram of $r$-band FWHM for the SAS2 
compared with the 3$\pi$ survey to date. 
The long tail to the  3$\pi$  observations is clear, although the modal 
seeing is only some 15 per cent higher than in SAS2. Note that we have not applied
any quality cut to the 3$\pi$ data, and many of the poorer seeing 
observations will be not be accepted for the final survey stacks. 
The other bands behave in similar fashion. The trend of worse seeing at 
shorter wavelengths has the sign expected from atmospheric seeing, but is
also believed to have a component due to the L2 corrector 
lens being slightly out of specification. In particular, the 
bluer bands suffer from a region of degraded seeing 
in the central half degree of the field.

\begin{figure}
\begin{center}
\centerline{\includegraphics[width=3.75in]{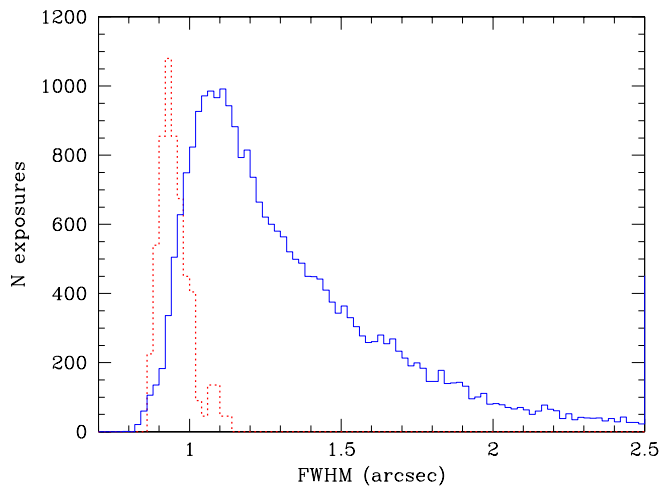}}
\caption{The distribution of mean $r$-band FWHM values for single exposures. Dashed line (red): the 124 SAS2 exposures, scaled by a factor of 45; Solid line (blue): the 26296 3$\pi$ exposures taken since the survey began.  The values are averages of a model fit to the stellar profile over the whole focal plane.}
\label{fig:rseeing}
\end{center}
\end{figure}

Fig. {\ref{fig:seeing} shows the distribution of FWHM measured on SAS2 
after the stacking process, again from model fits. The median  values are very
similar to those of the unstacked images in Table \ref{table:drm}, 
except possibly for the $y$ band, where there appears to have been a 
slight ($\sim10$ per cent) degradation in image quality. We are unable to explain
why this band should differ from any other, as all are treated identically
in the stacking process.

\begin{figure}
\begin{center}
\centerline{\includegraphics[width=3.75in]{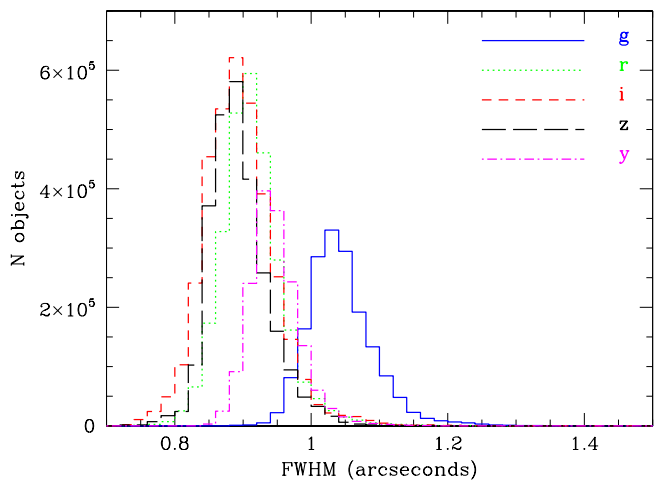}}
\caption{The distribution of model-fit FWHM values for all objects detected by the pipeline software on the SAS2 stacks. These agree well with the unstacked values (Table \ref{table:drm}), although the $y$-band has degraded by $\sim10$ per cent.}
\label{fig:seeing}
\end{center}
\end{figure}

\section{Data reduction and Analysis}
\label{sec:data}

\subsection{IPP processing pipeline}
\label{subsec:ipp}

The image processing pipeline reduction of PS1 images (IPP) is quite 
complex \citep[see e.g.][]{PS1_IPP} and it 
is not the aim of this paper to describe
these procedures in detail, but a 
brief overview is useful here.
Once detrended and astrometrically calibrated, the individual exposures are
resampled (`warped') on a fixed grid (a tangent plane projection) on the sky 
with a constant pixel size of $0.25$~arcsec (similar, but not identical to, 
the variable native pixel scale). 
This fixed grid is broken into a set of `skycells',
roughly $26$~arcmin across. Adjacent skycells are designed to have some 
overlap (to minimise issues with objects being cut by skycell boundaries, as 
photometry is performed independently on each skycell).
Depending on the orientation and alignment of the original exposure, 
as many as four OTAs may contribute to one skycell. Note that because 
of the gaps between the CCD cells and between the OTAs, $\sim 14$ per cent of the 
area of each of these skycells has no data. 

Skycells from different observations of the same field can then
be combined to produce deeper, stacked data. Outlier
rejection is applied to clean artifacts unique to individual images from 
the stack.
As with the individual exposures, the subsequent data analysis is 
carried out independently on each stacked skycell.

The main data analysis routine is {\sc psphot}. This constructs a 
model PSF for each skycell from high significance objects identified as 
point sources. The PSF is allowed to vary spatially over the field.
The model has functional form 

\begin{equation} \label{eq:ps1v1}
I=\frac{I_{0}}{1 + kr^2 + r^{3.33}} ,
\end{equation}

\noindent where $r$ represents a general, radial, elliptical 
coordinate ($r^2=\frac{x^2}{2\sigma_{xx}^2}+\frac{y^2}{2\sigma_{yy}^2}+\sigma_{xy}xy$). The model 
is force-fit to all objects (detected at greater than $5\sigma$ significance) to 
produce a PSF magnitude ({\tt CAL\_PSF\_MAG}).  Fig. \ref{fig:psf} shows how a 
typical example  of this profile, with $k=-0.123$ (the mean $k$ for SAS2 
is $-0.05$ with an rms scatter of $\pm0.1$), 
differs from a Gaussian with the same FWHM ($0.85$~arcsec) and total flux. 
Obviously there is more power in the wings of the PS1 profile, with only 
$\sim38$ per cent of the flux inside the FWHM. Even at 
20 pixel ($5$~arcsec) radius the PS1 profile is still a few percent short 
of recovering the total flux. 

Note that the IPP pipeline returns two quantities, {\tt PSF\_MAJOR} and {\tt PSF\_MINOR},
based on  $\sigma_{xx}$, $\sigma_{yy}$ and  $\sigma_{xy}$. 
For a circularly symmetric 
profile with $k=0$ these are by definition $0.5\times$ the FWHM of the 
profile (Figs. \ref{fig:rseeing} and \ref{fig:seeing} use this relation). 
For other $k$ this is only an approximation, although for the range of 
$k$ found in the real data deviations are only a few percent. For example, 
for the PS1 profile from Fig. \ref{fig:psf}, {\tt PSF\_MAJOR} $= 0.41$~arcsec, 
whereas the FWHM of the model is $0.425$~arcsec.

\begin{figure}
\begin{center}
\centerline{\includegraphics[width=3.45in]{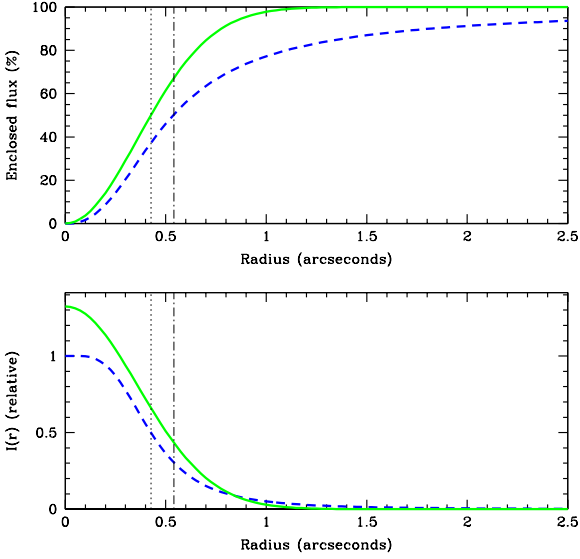}}
\caption{Top panel: the theoretical curves of growth for a Gaussian PSF (solid green line) and the model PS1 PSF described in the text with $k=-0.123$ 
(dashed blue line). Both profiles have the same FWHM and total intensity. 
The vertical dotted line indicates the radius which contains 50 per cent of the 
light for the Gaussian (half the FWHM), whilst the dot-dash line shows 
that for the PS1 profile.  Bottom panel: as above, but now for 
the intensity profiles, $I(r)$.}
\label{fig:psf}
\end{center}
\end{figure}

Obviously the PSF model is a useful measure for point
sources, but somewhat meaningless for most galaxies, so at the same time a 
Kron magnitude \citep{kron} is measured for the same objects, where the Kron 
flux is defined as the flux inside a circular radius of

\begin{equation} \label{eq:kron}
r_{\rm k} = 2.5 \times \frac{\sum r f(i)}{ \sum f(i)}
\end{equation}

\noindent where the sum is taken over a series of annuli, extending to 
large radii (in practice, IPP uses an iterative procedure, terminating the
summation at $6\times$ the radial moment found in the previous iteration), 
and  $f(i)$ is the light distribution curve (i.e. the radial intensity 
profile multiplied by the area of each annulus). 

The measurement of both PSF and Kron magnitudes requires a 
determination of the background sky. {\sc psphot} constructs 
a sky model on a regularly spaced grid (the spacing was 400 pixels for 
this analysis). To do this, a histogram of a randomly selected subset of the 
pixels in a box of side twice the grid spacing is constructed around each grid point (there is, therefore, some correlation between adjacent points). A robust
estimate of the peak is then made, which attempts to account for the skewed 
nature  of the histogram caused by the wings of bright objects.
To determine the sky at any position in the full-scale image, 
bilinear interpolation between the grid points is used. 

Other magnitudes (e.g. model fits to extended sources) will
be provided for the public release of the $3\pi$ data, but are still in the
testing and development stage, and so we do not consider them further in
this paper.

\subsection{Simulating PS1 data}
\label{subsec:simul}

The IPP pipeline is capable of adding (and recovering) fake stars when it 
performs photometric analysis of an image, in order to provide an estimate 
of depth. However, it does not simulate 
galaxies, so, in order to overcome this limitation and also to provide a full
independent investigation of SAS2, we generate our own fake stars 
and galaxies.  
A full description of how we construct these fakes will be given in 
Paper II. For this
paper it is only necessary to note that we use a PS1 profile for 
stars (Section \ref{subsec:ipp}), and a combination of exponential disks and 
de Vaucouleurs profiles for galaxies, and that 
we generate a realistic distribution of
galaxy properties, appropriate to the depth of the SAS2 images, using the 
mock galaxy catalogues of \citet{merson}, based on a $\Lambda$CDM 
cosmology, and the size distributions of \citet{shen}.

The simplest thing we can do is place these fake objects onto random 
backgrounds (with similar variance to real OTAs from single exposures) and 
try and recover them with {\sc psphot}.
\begin{figure}
\begin{center}
\centerline{\includegraphics[width=3.5in]{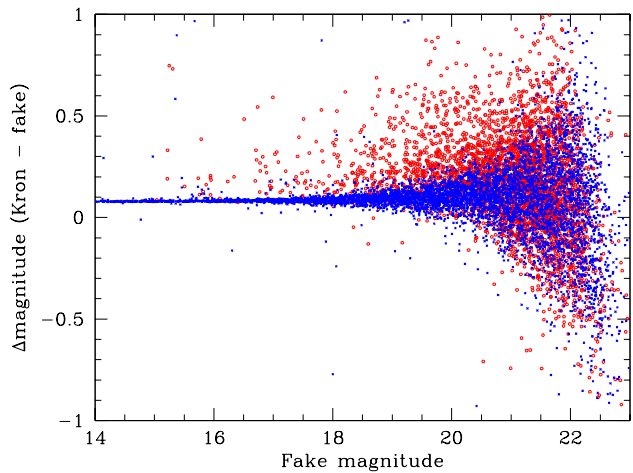}}
\caption{The difference between fake total magnitudes and Kron magnitudes returned from {\sc psphot} for stars (blue crosses) and galaxies (red circles). The 
objects have been placed on a random background, with variance and fluxes similar to that for real $r$-band SAS2 OTAs.}
\label{fig:fake_mags}
\end{center}
\end{figure}
Fig. \ref{fig:fake_mags}, where we compare Kron
magnitudes from {\sc psphot} with the true total input 
magnitude for each fake object, shows the results of such a simulation 
based on four simulated
SAS2 $r$-band OTAs, which have FWHM, mean variance, pixel masks, 
image scale and  flux calibration identical to their real counterparts 
(the corresponding real OTAs, XY21/22/31/32, were chosen as those which 
made up 
SAS2 warp, 454104, which has properties typical of the survey).
A numerical summary is given in Table \ref{tab:fakescatter}. 
An offset is always expected, as theoretically
a Kron magnitude only ever recovers a fixed fraction of the flux from
an object. For a Gaussian profile, with the Kron multiplier of 2.5 used here,
this fraction is $\sim0.99$ . For the PS1 PSF, equation \ref{eq:ps1v1}, it
is slightly dependent on $k$, but for the value of $k$ used here is 
about 95 per cent. The measured offset is $\sim0.06$~mag, 
which is very close to this prediction.
For galaxies the recovered fraction should be smaller, and dependent on the particular
profile of the object, being $\sim90$ per cent for a de Vaucouleurs $r^{1/4}$
profile. Table \ref{tab:fakescatter} suggests our fakes have an offset 
of $\sim0.2$~mag, which is slightly higher than expected.

\begin{table}
\caption{{\sc psphot} measured Kron magnitude offsets and scatter relative to the total fake magnitude for simulated objects placed on a random background. The fluxes and variance are equivalent to those for four real $r$-band SAS2 OTAs. These are the data shown in Fig. \ref{fig:fake_mags}.}
\begin{center}
\begin{tabular}{lcccc}
\hline
\hline
{\bf Magnitude} & \multicolumn{2}{c}{\bf $\Delta$~mag} (Kron - Fake) \\
(Fake)&(galaxies)&(stars)\\
\hline
$15-16$   &              & $0.06\pm0.03$ \\
$16-17$   &$0.18\pm0.14$ & $0.06\pm0.02$ \\
$17-18$   &$0.18\pm0.12$ & $0.07\pm0.02$ \\
$18-19$   &$0.20\pm0.13$ & $0.07\pm0.04$ \\
$19-20$   &$0.24\pm0.16$ & $0.08\pm0.06$ \\
$20-21$   &$0.19\pm0.17$ & $0.08\pm0.10$ \\
$21-22$   &$0.13\pm0.27$ & $0.07\pm0.22$ \\
\hline
\end{tabular}
\end{center}
\label{tab:fakescatter}
\end{table}

Fig. \ref{fig:de_chip} shows the recovered fraction 
(the `detection efficiency') of our fake stars and galaxies on the four 
simulated OTAs as a function of
their input magnitude. The error bars show the rms scatter between several 
different realisations. We also show the 
results of the built-in IPP pipeline detection efficiency measurements on the
the corresponding real OTAs. These are only
performed for stars, but there is good agreement between our
results and those from the IPP pipeline. 
The vertical line shows the position of the notional $5\sigma$ limit, where 
$\sigma$ is the measured noise inside an aperture of diameter equal to the 
FWHM of the PSF. 
This is seen to equate to a stellar detection efficiency of somewhere between 50 per cent and 60 per cent, a result we will return to in Section \ref{subsec:stack}.
The 50 per cent recovered fraction for galaxies occurs about $0.5$~mag brighter than for the stars, as might be expected given the more extended nature and hence lower surface brightness of the typical galaxy profile. The precise position of the
galaxy curve is, of course, dependent on how realistic our mock 
galaxy catalogues are (the galaxy profiles, morphological mix and redshift distribution all play a part) - we can be more confident for the stars, where the
only requirement is that we match the PS1 point spread function.

\begin{figure}
\begin{center}
\centerline{\includegraphics[width=3.5in]{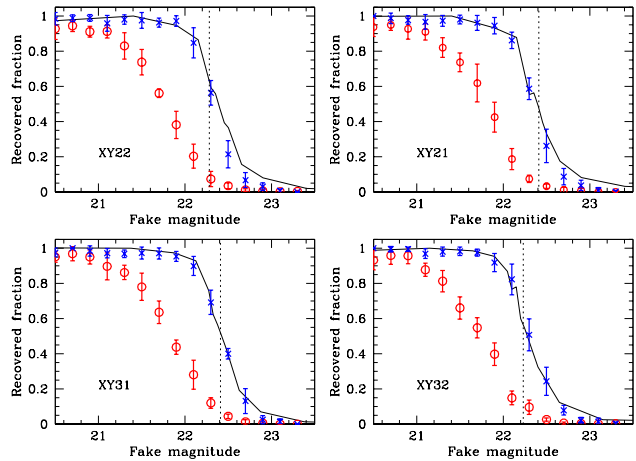}}
\caption{The fraction of fake point sources (blue stars) and galaxies (red circles) recovered by {\sc psphot} from four 
simulated  SAS2 OTAs, as a function of their simulated total magnitude. The dashed vertical line shows the magnitude of a point source which would have a flux
five times the background noise inside an aperture of diameter equal to the stellar FWHM. The solid lines show the recovered fraction of fake stars as measured 
by the IPP pipeline processing on the corresponding real OTAs.}
\label{fig:de_chip}
\end{center}
\end{figure}

As a final check we now place our fakes directly onto a real SAS2 stack. 
Here the variance is no longer constant across the field.
\begin{figure}
\begin{center}
\centerline{\includegraphics[width=3.5in]{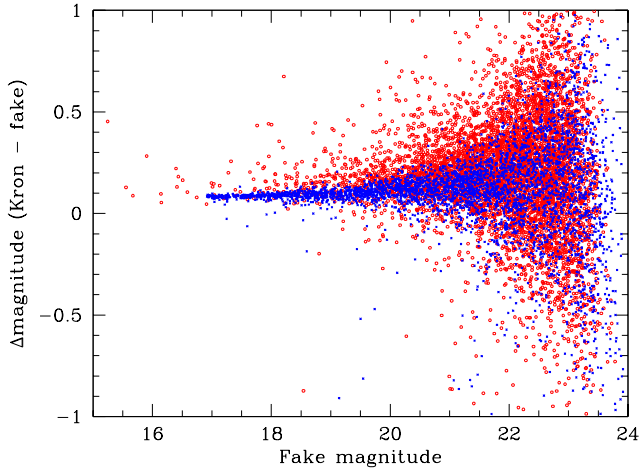}}
\caption{A comparison of the difference between input total magnitudes and measured Kron magnitudes for simulated stars (blue crosses) and galaxies (red circles) placed 
on a real SAS2 stack, as a function of their simulated magnitude.}
\label{fig:fakestars}
\end{center}
\end{figure}
Fig. \ref{fig:fakestars} shows the Kron magnitudes {\sc psphot} 
recovered 
from fake stars and galaxies placed on an $r$-band
SAS2 stack with typical seeing, compared with their simulated total 
magnitudes. Table \ref{tab:fakestackscatter} summarises the comparison numerically.
The magnitudes of the stars are drawn from a realistic power-law 
distribution, whilst the galaxies are taken from the mock catalogues
discussed earlier.  On the whole, {\sc psphot} does
a good job, and gives similar results to those for fakes on a random
fake background (Table \ref{tab:fakescatter}). Remember, as discussed earlier 
in this section, we expect that Kron magnitudes will always be 
slightly fainter than the true total magnitude, so the offsets displayed in
Fig. \ref{fig:fakestars} are expected. In fact, for galaxies, the offsets
are nearer the theoretical expectation than they were on the fake backgrounds.
There is a slight trend for objects to be measured systematically too faint 
as the true magnitude becomes fainter (by about $0.05$ magnitudes over 
the range $17<r<22$). We will return to this issue in Section \ref{sec:sdsspixels}.

\begin{table}
\caption{{\sc psphot} measured Kron magnitude offsets and scatter relative to the total fake magnitude for $25000$ simulated stars and $38000$ simulated galaxies placed on real SAS2 $r$-band stack. These data are shown in Fig. \ref{fig:fakestars}.}
\begin{center}
\begin{tabular}{lcccc}
\hline
\hline
{\bf Magnitude} & \multicolumn{2}{c}{\bf $\Delta$~mag} (Kron - Fake) \\
(Fake)&(galaxies)&(stars)\\
\hline
$17-18$   &$0.13\pm0.09$ & $0.06\pm0.04$ \\
$18-19$   &$0.13\pm0.11$ & $0.07\pm0.05$ \\
$19-20$   &$0.14\pm0.12$ & $0.08\pm0.06$ \\
$20-21$   &$0.17\pm0.17$ & $0.09\pm0.09$ \\
$21-22$   &$0.20\pm0.23$ & $0.10\pm0.15$ \\
$22-23$   &$0.19\pm0.35$ & $0.12\pm0.29$ \\
\hline
\end{tabular}
\end{center}
\label{tab:fakestackscatter}
\end{table}

\subsection{The effects of warping and stacking}
\label{subsec:warp}
As described above, the PS1 data go through both warping and stacking
stages before reaching the final data product, either of which might 
potentially loose depth. The warping stage, in particular, convolves the 
data on small scales using a Lanczos3 kernel.
The effect can be seen in Fig. \ref{fig:power} which shows the pixel 
power spectrum as a function of wavenumber for an SAS2 OTA, an SAS2 warp, 
and, for comparison, a single 
SDSS field. Detected objects have been masked out before the power spectrum 
was computed, and for comparison purposes we have renormalised each 
so that they overlap at $\sim1.5$~arcsec.
The expectation from Poisson noise from sky (and read noise
from the detector) would be a flat power spectrum on all scales. 
As expected, for the SAS2 warp we see a sharp downturn in power for scales 
above $k\sim0.9$ arcsec$^{-1}$, which is due to the smoothing introduced 
by the warping 
process. The OTA and SDSS power spectra are similar at most scales - 
the SDSS data cuts off at a smaller wavenumber due to its larger pixel size 
($0.4$~arcsec compared with $0.256$~arcsec for the PS1 OTA). Both PS1 power spectra do show 
small spikes at log k $\sim0.58$ and $\sim0.87$. We believe these are related to
problems with the variable bias structure discussed further in Section 6.

\begin{figure}
\begin{center}
\centerline{\includegraphics[width=3.5in]{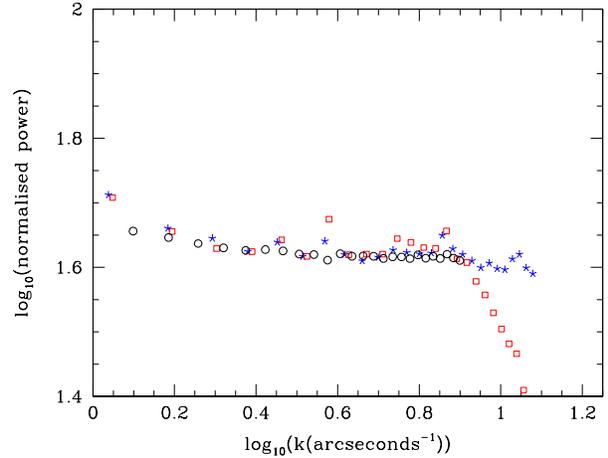}}
\caption{The pixel flux power spectrum as a function of angular wavenumber for an SAS2 chip (blue stars), and SAS2
warp (red squares) and an SDSS tile (black circles). An arbitrary renormalisation has been applied to bring the
three into agreement on scales of $\sim1.5$~arcsec. The loss of power on small scales due to the smoothing effect of the warping process is clearly visible.}
\label{fig:power}
\end{center}
\end{figure}

To investigate the consequences of this smoothing, we consider the recovery 
of fake stars placed on PS1 OTAs which then go through the warping
process. Note that we measure the recovered fraction as a 
function of simulated magnitude, and we take no account of 
the actual measured magnitudes of the fakes 
(although, in general, at the 50 per cent recovery magnitude, the 
offset between measured and input magnitude is $<0.1$~mag). 

\begin{figure}
\begin{center}
\centerline{\includegraphics[width=3.5in]{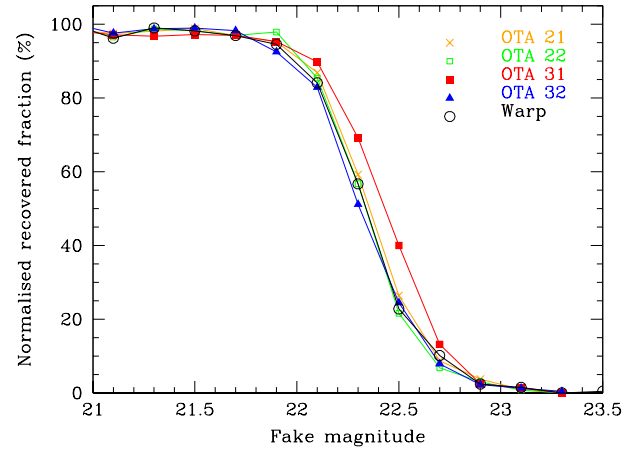}}
\caption{The percentage of fake stars recovered by {\sc psphot}, as a 
function of input magnitude, for the four simulated OTAs, and the warp
created from them. These are the same OTAs as were simulated 
in Fig. \ref{fig:de_chip}. Any loss in depth caused by the warping process 
would result in the warp line moving to the left of the relationships 
for the OTAs.}
\label{fig:chipwarp}
\end{center}
\end{figure}

Fig. \ref{fig:chipwarp} shows the detection efficiency curves for our four 
simulated OTAs (these 
are the same OTAs as used in Fig. \ref{fig:de_chip}), and that for the warp 
created from these OTAs using the IPP pipeline.
Note that due to the different fraction of masked pixels 
in each, it is necessary to normalise the curves to produce a 100 per cent
recovery at bright magnitudes in order to intercompare.
In an ideal world the warping process would not lose any depth, and all 
five curves would overlap. In practice, not all the OTAs have exactly
the same noise, and they do not all contribute equal areas to the warp (OTAs
XY31 and XY32 contribute slightly more area than XY21 and XY22), 
so it is not easy to judge. OTA XY31 is slightly deeper than the warp, but 
the other OTAs are not, so it seems unlikely
that the warping process results in any loss of depth 
$\ge0.05$~mag.

Having checked the warping procedure we now investigate the stacking
procedure. The same fake stars and galaxies are added to each of the 
14 warps which  make up the SAS2 r-band stack 1034502 (one of which is 
warp 454105 which was used in the warping test). These warps are then 
put through the IPP stacking routine {\sc ppstack}, and  {\sc psphot} run on
the resulting stack. The fraction of the fakes recovered as a function of 
their fake magnitude can then be determined. This is a 
slightly more rigorous test
then relying on the pipeline detection efficiency routine, as this puts 
independent fakes (i.e. at different locations on the sky) on each warp, and 
performs forced photometry at these locations. Obviously these 
cannot then be followed through the stacking process.

Fig. \ref{fig:warps} show the results for stars. The warps, of course, 
are not identical, and it has been
necessary to normalise each of the warp curves slightly
to agree at bright magnitudes, due to the variation in masked fraction. There
is also some natural variation in depth - the hatched area shows the spread in recovered fraction between them.
As the mean number of warps per pixel in the resulting stack is measured to be 8.8, we would expect the stack to 
be $2.5\times \log{(\sqrt{8.8})} = 1.18$ magnitudes fainter. This offset 
is shown by the arrow on Fig. \ref{fig:warps}, and is seen to be in 
excellent agreement with the data, so we are confident the stacking process is behaving as expected.

Fig. \ref{fig:stacks} compares the recovery of stars and galaxies on the stack.
Our fake galaxies have a 50 per cent completeness about $0.4$~mag brighter than 
the stars.  This is very slightly smaller than on the chips 
(Fig. \ref{fig:de_chip}), which presumably represents the fact that the galaxies profiles become more seeing-dominated, and hence appear star-like, at the deeper limit of the stack. 
Also shown is the result of running the pipeline detection efficiency
routine on the equivalent real SAS2 stack. This puts fake stars directly onto the stack, rather than following 
them through the stacking process, but nevertheless the results are very similar.

\begin{figure}
\begin{center}
\centerline{\includegraphics[width=3.75in]{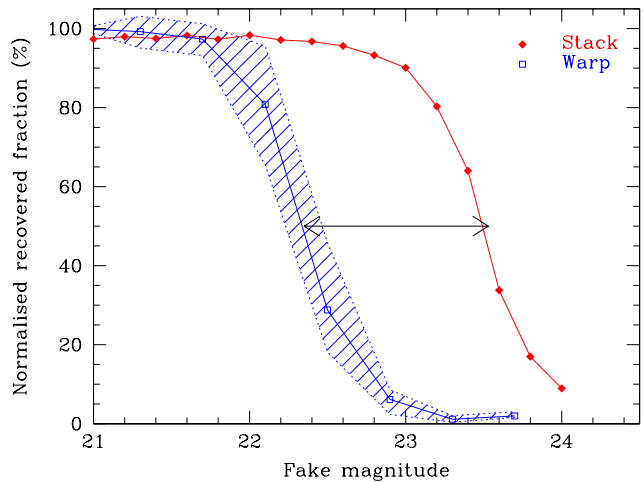}}
\caption{The mean recovered fraction of our fake stars over the 14 warps which
make up our $r$-band stack ((blue, open squares), and the fraction on the stack itself (red solid diamonds), plotted as a function of their 
true total magnitude. The double-headed arrow shows the increased depth expected from the stacking process. The hatched zone shows the range covered by 
each of the 14 separate warps. Each warp has been renormalised to give a recovered fraction of 1 at $r=20.5$. }
\label{fig:warps}
\end{center}
\end{figure}

\begin{figure}
\begin{center}
\centerline{\includegraphics[width=3.5in]{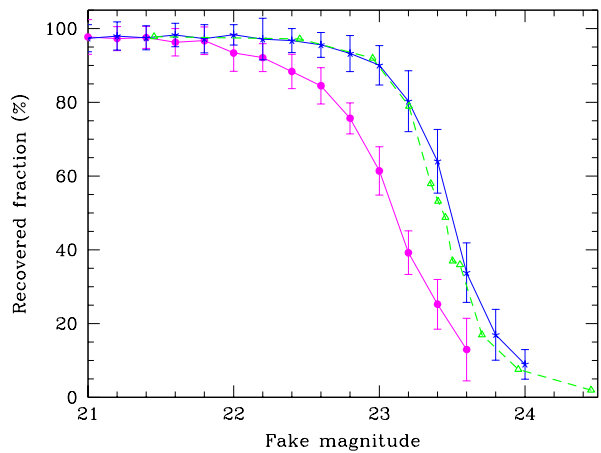}}
\caption{The recovered fraction of our fake stars (blue asterisks) and 
galaxies (magenta dots) on our test stack, as a function of 
their true magnitude. Error bars indicate the rms variation between several 
realisations of the fake warps which go into this stack. Also shown (green triangles, dashed line) 
are the fake star results from the pipeline for the equivalent PS1 stack.}
\label{fig:stacks}
\end{center}
\end{figure}

\subsection{Predicting the noise}
\label{subsec:noise}

We now wish to ask whether the depths of the SAS2 stacks are in line
with those expected given the known properties of the camera and the 
conditions under which the images were taken.
In this section we address whether the measured depth 
of SAS2 single exposures is in line with the prediction based on the 
measured sky and expected read-noise of the system. We define the noise
per pixel of a single exposure  as

\begin{equation}\label{eq:noise}
\sigma = \sqrt{(s + d + rn^2)}
\end{equation}

\noindent where {\it s} is the flux recorded from the background sky, {\it d} the dark 
current accumulated during the exposure and {\it rn} the read-noise of the 
camera. All quantities are measured in electrons. We assume a fixed gain 
of $1$~e$^-$~ADU$^{-1}$ and readnoise of $\sim5.5$~e$^-$
(these are typical of the values recorded in the GPC1 camera
image file headers for each CCD cell).
The dark current is taken to be  $\sim0.2$~e$^-$~s$^{-1}$ \citep{camera}.
In practice, all three quantities may vary by a few percent between OTAs, but
we do not believe this will have any significant effect upon our analysis.
The actual sky and the sky variance on each OTA
are measured as part of the IPP data reduction  process (after detrending).

\begin{table}
\caption{Predicted and measured noise levels per exposure (in ADU per pixel,
assuming a gain of $1$~e$^-$~ADU$^{-1}$), for the subset of the SAS2 which has SDSS
Stripe 82 coverage.
The errors come from the variation between
exposures and do not indicate the accuracy to which the individual quantities
can be measured on a single exposure.}
\begin{center}
\begin{tabular}{lccl}
\hline
\hline
{\bf Band} & {\bf Measured} & {\bf Predicted} & {\bf Sky ADU}\\
&{\bf $\sigma$}&{\bf $\sigma$}&\\
\hline
\gps  &   $7.92\pm0.04$ & $8.18\pm0.03$  & $28.1\pm0.5$\\
\rps  &  $10.11\pm0.04$ & $10.36\pm0.04$ & $69.0\pm2.7$ \\
\ips  &  $12.48\pm0.11$ & $13.09\pm0.10$ & $132.2\pm2.6$\\
\zps  &  $12.97\pm0.11$ & $13.03\pm0.09$ & $133.6\pm3.8$ \\
\yps  &  $13.96\pm0.10$ & $14.17\pm0.11$ & $161.9\pm3.0$\\
\hline
\end{tabular}
\end{center}
\label{table:sigma}
\end{table}

Table \ref{table:sigma} shows the results for all the exposures used in the
SAS2 which have Stripe 82 coverage (about 160 out of 600 skycells), separated
by filter. The predicted values are slightly higher (up to 5 per cent) than those
observed, which may suggest that the read noise and/or dark current have
been slightly  overestimated. However, given the uncertainties, the
results are reasonably consistent, and we now go on to predict the variance
expected on the stacks.

\subsection{The depth of the stacks}
\label{subsec:stack}

Having seen in Sections \ref{subsec:warp} and \ref{subsec:noise} that 
the warping and stacking processes are reasonably well behaved, and the
noise levels in single exposures are much as expected,
we turn our attention to
the depth of the stacked SAS2 exposures. We investigate two measures of 
depth: (1) the magnitude at which 50 per cent of fake stars input onto the stacks
are recovered, and (2) the magnitude at which the differential 
number-magnitude count of all objects peaks.  We would like to compare
these {\sc psphot} results with the expected $5\sigma$ limit for 
a point source inside a circular aperture of diameter equal to the FWHM.
To do this it is necessary to modify the equation \ref{eq:noise} to allow 
for the number of warps going into each pixel in the stacked skycell. 
 
\begin{equation}
\sigma = \sqrt{((s+d+rn^2) \times coverage)}
\end{equation}

\noindent where the coverage is the average of the  number of warps 
contributing to each pixel.
Due to the construction of the camera, this coverage factor is not simply 
equal to the number of input warps (ideally 12 for the SAS2 and the final 
$3\pi$ surveys, although this number varies slightly due to the exact 
pattern of exposures on the sky), as $\sim14$ per cent of the sky is lost on each
warp to the gaps between the detectors 
(see Section \ref{subsec:ipp}). In fact, once cosmetic masking, particularly
of defective CCD cells, has been taken into account the true losses can be 
significantly higher. As one of the IPP products is a coverage map 
for each stacked skycell, it is easy to determine the
actual value, which turns out to be around $8.9\pm0.9$ warps per pixel 
for a skycell 
in the central region of SAS2, where we have full coverage. This implies an
average masked fraction of around 25 per cent per warp (actually, some losses may
come from outlier rejection during the stacking process itself, so individual
warps may not be as bad as this).

To plot the $5\sigma$ limit, we assume the idealised case of a Gaussian 
stellar profile, so the total flux is $2.0$ times that inside a diameter
equal to the FWHM. Note, however, that the PS1 stellar profile 
is not Gaussian, so these $5\sigma$ limits just act as a fiducial marker, 
and to determine the expected absolute offset to {\sc psphot} 
magnitudes requires the simulations in Section \ref{subsec:simul}.

\begin{figure}
\begin{center}
\centerline{\includegraphics[width=3.5in]{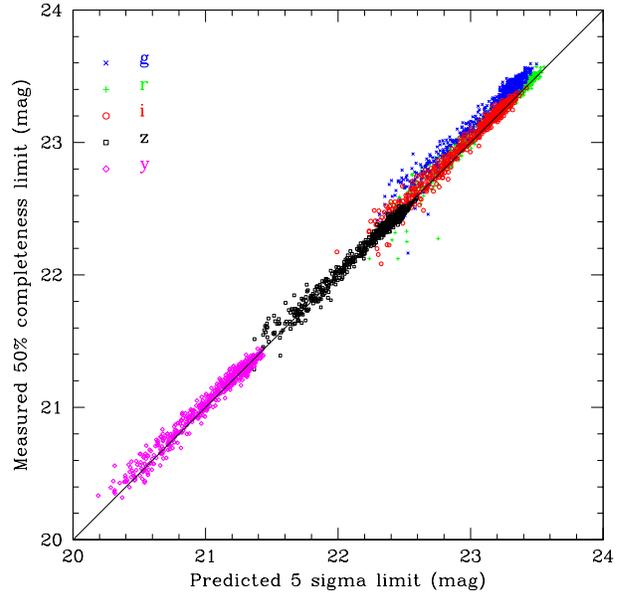}}
\caption{The simulated magnitude at which 50 per cent of fake stars injected into all the SAS2 stacks are recovered, compared with the predicted $5\sigma$ depth based on exposure time, the measured seeing, sky background and zero point, and assuming a Gaussian PSF.}
\label{fig:limit50}
\end{center}
\end{figure}

There are some limitations to our analysis:
\begin{itemize}
\item{The measured FWHM from {\sc psphot} are based on a 
PS1 stellar image profile, although we used them to calculate the Gaussian
$5\sigma$ limits. We have shown in Fig.~\ref{fig:psf} that the two
are very similar. However, this uncertainty needs to be
borne in mind when comparing with surveys from other telescopes. 
As a rough guide, a $0.1$~arcsec difference in FWHM would make a 
$\sim0.1$~mag difference to our predicted $M_{5\sigma}$ limiting magnitude.}

\item{As noted in Section \ref{subsec:noise} the fixed values we assume 
for readnoise, dark current and gain may, in practise, vary
slightly from OTA to OTA. However, we believe the effect of these
assumptions to be small (the exposures in the redder bands are sky-noise 
limited).}

\item{We use values for the sky and sky variance which are averaged over 
the whole exposure, rather than for the particular OTAs which contribute to 
a skycell. Again we believe the effect of this to be small.}

\item{Although we scale the variances by coverage factor, we do not account 
for the slight smoothing caused by the warping process. So our $5\sigma$ 
limits apply to an idealized, unwarped stack. }

\end{itemize}

Bearing all this in mind, we now compare our predictions to the 
detection efficiency limits measured by the IPP pipeline. 
Fig. \ref{fig:limit50} shows the simulated  magnitude at which 50 per cent 
of the fake stars 
injected into the stacks are recovered as a function of the predicted 
$5\sigma$ limit, 
for all five filters. The fake stars have profiles and FWHM equivalent to
those measured for real stars on each stack, and vary spatially in the same
fashion.
The main point to take from this figure is that there is a one-to-one 
relationship between the two quantities which is followed by all the bands, 
and for different stack exposures times (the spread along the diagonal 
direction
within each band is due mainly to differing effective stack exposures 
caused by the reduced coverage towards the edges of SAS2). This suggests
the stacked data are all well behaved.

Due to the uncertainties already mentioned, the absolute offset between the 
two axes (which is close to zero) is more difficult to interpret. The 
points do lie reasonably close to the expectations of our simulations 
in Section \ref{subsec:simul}, where we showed that the $5\sigma$ limit 
corresponded to a 50-60 per cent recovery fraction for fake stars. 
We also note that in an ideal case, cutting a sample at 
a magnitude equivalent to n-$\sigma$ should always result in 50 per cent of 
objects 
at that magnitude being
detected. In reality the situation is more complex: (a) errors, such as 
those caused by the determination of the sky background, might not
scatter equal numbers brightward and faintward of their true magnitude; (b)
the process of detecting objects may depend on factors only indirectly
related to the random noise; (c) as is the case here, the estimate of 
n-$\sigma$ might not perfectly match the definition used to limit the sample.
However, our tests seem to show that the effects of these uncertainties are
quite small, and that we do recover around 50\% of objects at our idealised 
$5\sigma$ limit.

The detection efficiency limits are, of course, derived from fake
objects put down with stellar profiles. In the real world, except near the 
galactic plane, most objects at these magnitude limits in the $3\pi$ survey
are going to be galaxies, and so would be expected to have shallower 
detection limits than for point sources.  It is also more appropriate to
use Kron magnitudes than PSF magnitudes. So we now show in 
Fig. \ref{fig:numdepth} the measured Kron magnitude at the which the 
differential number 
counts of all objects on each stack peaks as a function of the $5\sigma$ limit 
(so a comparison can be made with Fig. \ref{fig:limit50}). 
As expected the turn-over Kron magnitudes are considerably brighter 
(by about $\sim0.6$~mag) than the 50 per cent PSF magnitude limits. This comes
partly from the lower detection efficiency for galaxies (see, e.g., Fig. \ref{fig:stacks})
and partly because, in general, the count peak occurs at a magnitude somewhat brighter 
than the 50 per cent limit.
The much larger scatter is probably due to the uncertainty in measuring the
peak. Although the downturn in the counts is very sharp as a function of PSF 
magnitudes (as the sample is limited in PSF magnitude), the corresponding 
turn-over is much shallower as a function of Kron magnitude, due to the 
intrinsic spread in $\Delta(m_{\rm Kron}-m_{\rm PSF})$.

\begin{figure}
\begin{center}
\centerline{\includegraphics[width=3.5in]{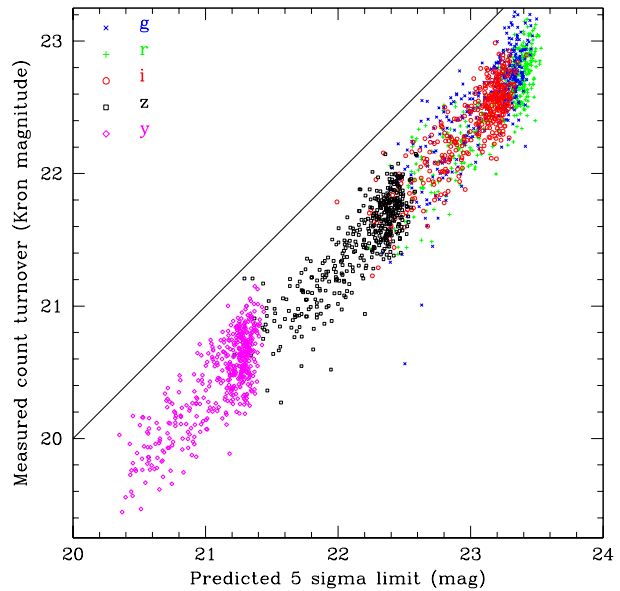}}
\caption{The Kron magnitude at which the differential object counts on the SAS2 stacks peak compared with the predicted $5\sigma$ depth based on exposure time, seeing and measured sky. The line of zero offset is provided only as a guide.}
\label{fig:numdepth}
\end{center}
\end{figure}

Fig. \ref{fig:spatialpercent} shows how the average depth per skycell 
(in this case the 50 per cent 
completeness magnitude in the \rps\ band) varies with position across 
the SAS2 field. By design, the central $8\times8$ deg region is very 
uniform - as expected the depth decreases at the edges of the field 
where coverage is less complete. In Paper II we will investigate how
the depth varies at higher spatial resolution than that of a single skycell.

\begin{figure}
\begin{center}
\centerline{\includegraphics[width=3.5in]{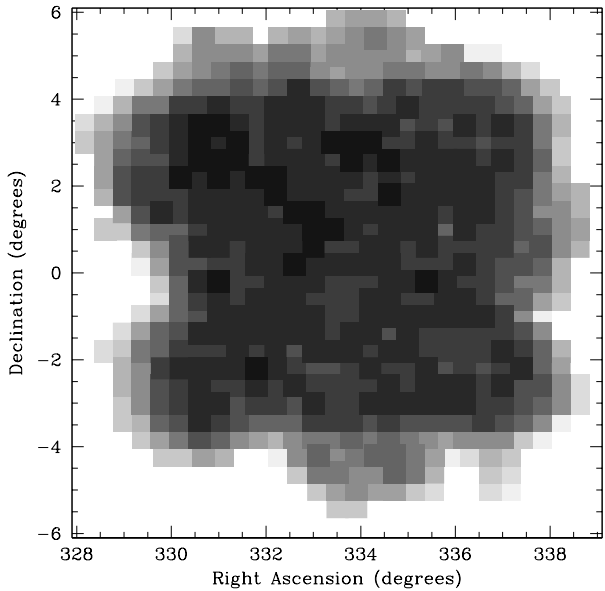}}
\caption{The spatial variation of the \rps\ 50 per cent completeness limit 
across SAS2 (skycell by skycell). The greyscale runs from 22.5 to 23.5 
in steps of 0.1~mag, light-grey to black.}
\label{fig:spatialpercent}
\end{center}
\end{figure}

\section{Running {\sc psphot} on SDSS pixels}
\label{sec:sdsspixels}
Our next test is to run {\sc psphot} on ten $r$-band fields from 
the SDSS DR8 release which are covered by SAS2, namely 
frame-r-004192-5-0171/72/73/74/75/87/88/89/90/91.
The FWHM on these fields is about $0.9$~arcsec. 
We use the same parameters as used for the PS1
data, with the proviso that {\sc psphot} uses a different PSF 
model for SDSS data than for PS1 data; specifically, $I(r) \propto (1 + kr^2 + kr^{4.5})^{-1}$. We take the photometric zero-point from the SDSS image 
headers. These fields cover a similar area to PS1 SAS2 skycells 1405.012/13/14.
There are about 7500 objects in the DR8 catalogue to the SDSS $5\sigma$ limit.

\begin{figure}
\begin{center}
\centerline{\includegraphics[width=3.75in]{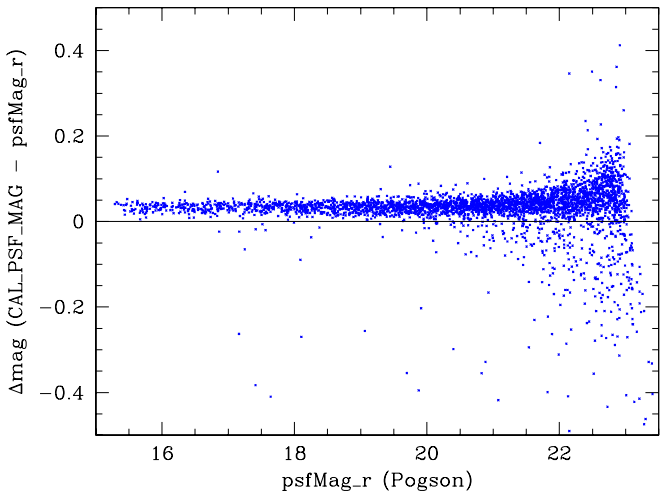}}
\caption{A comparison of PSF magnitudes measured by {\sc psphot} on the 10 $r$-band SDSS fields and the DR8 catalogue psfMag magnitudes (corrected to be Pogson) for the same 
objects. Only objects classed as stellar in DR8 ({\tt type}$=6$) are shown.}
\label{fig:sdsstile_psf}
\end{center}
\end{figure}

\begin{table}
\caption{PSF magnitude offsets and $rms$ scatter relative to SDSS DR8 Pogson-corrected psfMags, for our {\sc psphot} reduction of the SDSS fields
and our SAS2 data on the same region. Only objects classified as stellar in DR8 are used.These are the data shown in Figs \ref{fig:sdsstile_psf}. Offsets are in the sense PS1 - DR8.}
\begin{center}
\begin{tabular}{lc}
\hline
\hline
{\bf DR8 mag.} & {\bf $\Delta$ magnitude}\\
($r$ Pogson)& ({\sc psphot} - DR8)\\
\hline
$<18$     & $0.03\pm0.01$ \\
$18-20$   & $0.03\pm0.02$ \\
$20-21$   & $0.03\pm0.08$ \\
$21-22$   & $0.03\pm0.04$ \\
$22-22.5$ & $0.04\pm0.07$ \\
$22.5-23$  & $0.06\pm0.08$ \\
\hline

\end{tabular}
\end{center}
\label{tab:tilescatter}
\end{table}

Fig. \ref{fig:sdsstile_psf} shows the comparison between our reduction of
the SDSS fields and the original DR8 catalogue magnitudes (converted to
Pogson magnitudes from Luptitudes - this correction only has a significant effect faintward of $r\sim22$, amounting to $0.04$ mag at 
$r=23$), for PSF-fitted magnitudes (in both cases) for objects 
classed as stellar in DR8 ({\tt type}$=6$).
Duplicate objects in areas of overlap between the SDSS fields have been 
removed. 
The results are summarised numerically in Table \ref{tab:tilescatter}. 
The agreement is good over the whole magnitude range from
$r =$ 15 -- 23, with a scatter of only $\le0.01$~mag brightward of $r\sim20$. 
There is a small offset (in the sense {\sc psphot}-DR8) of 
$\sim0.03$~mag, suggesting the two fitting techniques measure slightly
different fluxes for the same objects. Aperture photometry on the images 
favours the SDSS values, so this is probably related to the 
amount of flux in the wings of the PSF model used by {\sc psphot}. 
If we force {\sc psphot} to adopt the usual PS1 model described in 
Section \ref{subsec:ipp} which has more extended 
wings, we find this offset disappears, but at the expense of a 50 per cent 
increase in the rms scatter. Note that, in practice, if we followed 
the full calibration procedure used for PS1 the zero-point would 
change to take out the offset anyway (although this might induce an offset 
in the opposite direction in extended source photometry).


Apparent visually, faintward of $r\sim22$, is a very small scale
error, with the offset increasing to about $+0.06$~mag 
by $r\sim23$. One possible
explanation for this scale error would be if {\sc psphot} measured a
slightly higher sky value than SDSS, but it could also be due to differences
in the PSF profile used.

\begin{figure}
\begin{center}
\centerline{\includegraphics[width=3.75in]{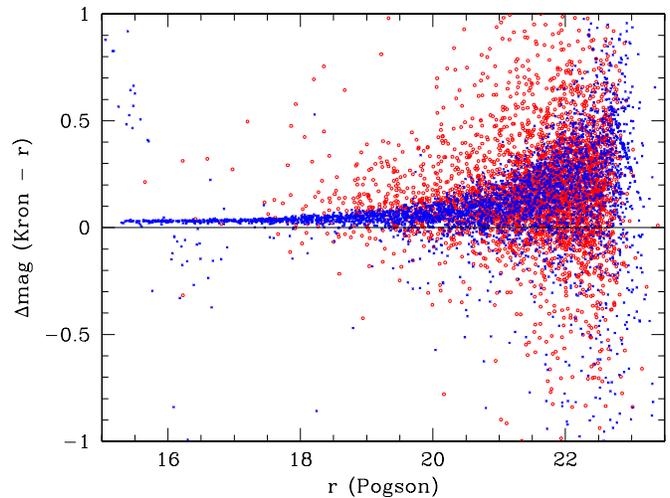}}
\caption{A comparison of Kron magnitudes measured by {\sc psphot} on the 10 SDSS fields, 
and the DR8 catalogue Pogson-corrected modelMag values for the same objects.
Blue crosses: objects classed as stellar in DR8 ({\tt type}$=6$); red circles - DR8 galaxies ({\tt type}$=3$).}
\label{fig:sdsstile_kron}
\end{center}
\end{figure}

Fig. \ref{fig:sdsstile_kron} shows the corresponding plot for Kron magnitudes compared with SDSS model magnitudes.
We now include galaxies as well as stars. Table \ref{tab:scatter} summarises the results.
It is clear that the Kron magnitudes are not behaving as well as the PSF magnitudes, especially for 
stars, which now show an obvious scale error. This is in the sense that the
magnitudes measured by {\sc psphot} become
systematically too faint at fainter SDSS magnitudes, with the offset
rising from $0.02$ to $0.18$ magnitudes.  
Puzzlingly, neither Table \ref{tab:fakescatter},
which shows the results of running {\sc psphot} on fake stars on
fake OTAs, nor Table \ref{tab:fakestackscatter}, which shows the same
for real stacks, show this problem to anything like this degree. And
although in Section \ref{sec:surveys} we will see that the effect
is present in our comparison between our SAS2 data and Stripe 82, again
it is at a much lower level.

We believe the most likely explanation for such a scale error is the 
underestimation of the Kron radii for faint objects due to the poor 
signal-to-noise in the outer regions of the profile. Indeed, the measured
Kron radius for stellar objects drops by about 25 per cent in the range 
$16<r<21$. Another possibility is that {\sc psphot} is overestimating
the sky background. However, tests with our fake objects suggest that, if
anything, {\sc psphot} slightly {\it underestimates} sky. 
It is unlikely that the problem lies in
the DR8 modelMags, as a direct comparison between the PS1 PSF and 
Kron magnitudes still shows the problem. One final possibility is that
many of the fainter objects classified as stars in DR8 are really
galaxies, for which Kron magnitudes, as we have already noted,  
recover a smaller fraction of the total flux than they do for stars. This 
might be of consequence for the faintest bin in Table \ref{tab:scatter}, but
at brighter magnitudes it is fairly unambiguous what is a star and what is 
a galaxy. None of these potential explanations, however, offer an insight as 
to why the effect is so much worse on the SDSS fields.

\begin{table}
\caption{Kron magnitude offsets and scatter relative to SDSS DR8 Pogson-corrected modelMags, for our {\sc psphot} reduction of the SDSS fields 
and our SAS2 data on the same region. These are the data shown in Fig. \ref{fig:sdsstile_kron}. Offsets are in the sense PS1 - DR8.}
\begin{center}
\begin{tabular}{lcccc}
\hline
\hline
{\bf DR8 mag.} & \multicolumn{2}{c}{\bf $\Delta$ mag.} ({\sc psphot} - SDSS) \\
($r$ Pogson)&(galaxies)&(stars)\\
\hline
$<18$     &              & $0.02\pm0.01$ \\
$18-20$   &$0.11\pm0.16$ & $0.04\pm0.04$ \\
$20-21$   &$0.16\pm0.26$ & $0.06\pm0.08$ \\
$21-22$   &$0.18\pm0.26$ & $0.12\pm0.18$ \\
$22-22.5$ &$0.17\pm0.26$ & $0.18\pm0.30$ \\
\hline
\end{tabular}
\end{center}
\label{tab:scatter}
\end{table}

We now turn our interest to the depth of the {\sc psphot} 
reduction compared
to that of the DR8 catalogues. Given that the two reductions were performed
on the same SDSS pixels, we would expect the counts to be very similar.
We take the deeper SDSS Stripe 82 catalogue (which covers the same region of
sky) as the `truth', and match our detections, and those of DR8, to this, 
using a circular match radius of $1.0$~arcsec\ (the rms scatter in separation of
our matched objects is only $\pm0.1$~arcsec\ in both RA and Dec., so this
match radius is more than adequate).
We restrict the DR8 catalogue to those objects with the $r$-band BINNED\_1 flag
set, i.e. they are $5\sigma$ detections on the $r$-band frame, as this is
the default limit of the {\sc psphot} code (in practice, this made
virtually no difference to our results).
Fig. \ref{fig:sdsstile_count} shows the differential $r$-band number counts of
matching objects, as a function of PSF-fitted magnitude.

\begin{figure}
\begin{center}
\centerline{\includegraphics[width=3.75in]{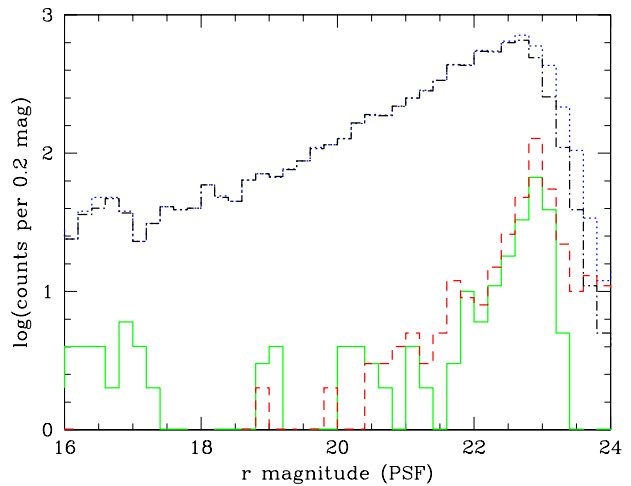}}
\caption{Differential number counts of objects on our test SDSS fields. The dot-dash (black) line shows the counts from our {\sc psphot} reduction which have matching counterparts in the SDSS Stripe 82 catalogue. The dotted (blue) line shows the same matched count but now for the DR8 catalogue. Both are plotted as a function of Stripe 82 r\_psfMag. The solid (green) line shows number of {\sc psphot} objects which do not have a match in the Stripe 82 catalogue, as a function of {\tt CAL\_PSF\_MAG}, whilst the dashed (red) line shows the unmatched DR8 counts as a function of DR8 r\_psfMag. The offsets between Stripe 82 r\_psfMag, {\tt CAL\_PSF\_MAG} and DR8 r\_psfMag are negligible.}
\label{fig:sdsstile_count}
\end{center}
\end{figure}

The two data-sets show very similar counts, both dropping sharply faintward
of the same magnitude limit ($r\sim23$) to within 0.1~mag.
Fig. \ref{fig:sdsstile_count} also shows the
number of unmatched objects, which are presumably false detections. 
Again these are very similar - if anything,
the {\sc psphot} reduction does slightly better. We deduce from this that
{\sc psphot} is performing at least as well as the SDSS software.

\section{Comparison with other Surveys}
\label{sec:surveys}
Having examined the internal consistency of the PS1 data, we now 
compare the $g$, $r$, $i$ and $z$ counts of objects in the stacked SAS2 with those from the SDSS DR8 
and co-added Stripe 82 catalogues. It is necessary to remove areas, mainly around
bright objects, where there are holes in the Stripe 82 catalogue. After doing this,
we are left with an area in common between all three surveys of $\sim16$ sq.deg.. 
We restrict the SDSS objects to those with the {\tt BINNED1} flag set to TRUE
(a $5\sigma$ detection) for the band in question ($g,r,i$ and $z$).

\begin{figure*}
\begin{center}
\centerline{\includegraphics[width=7in]{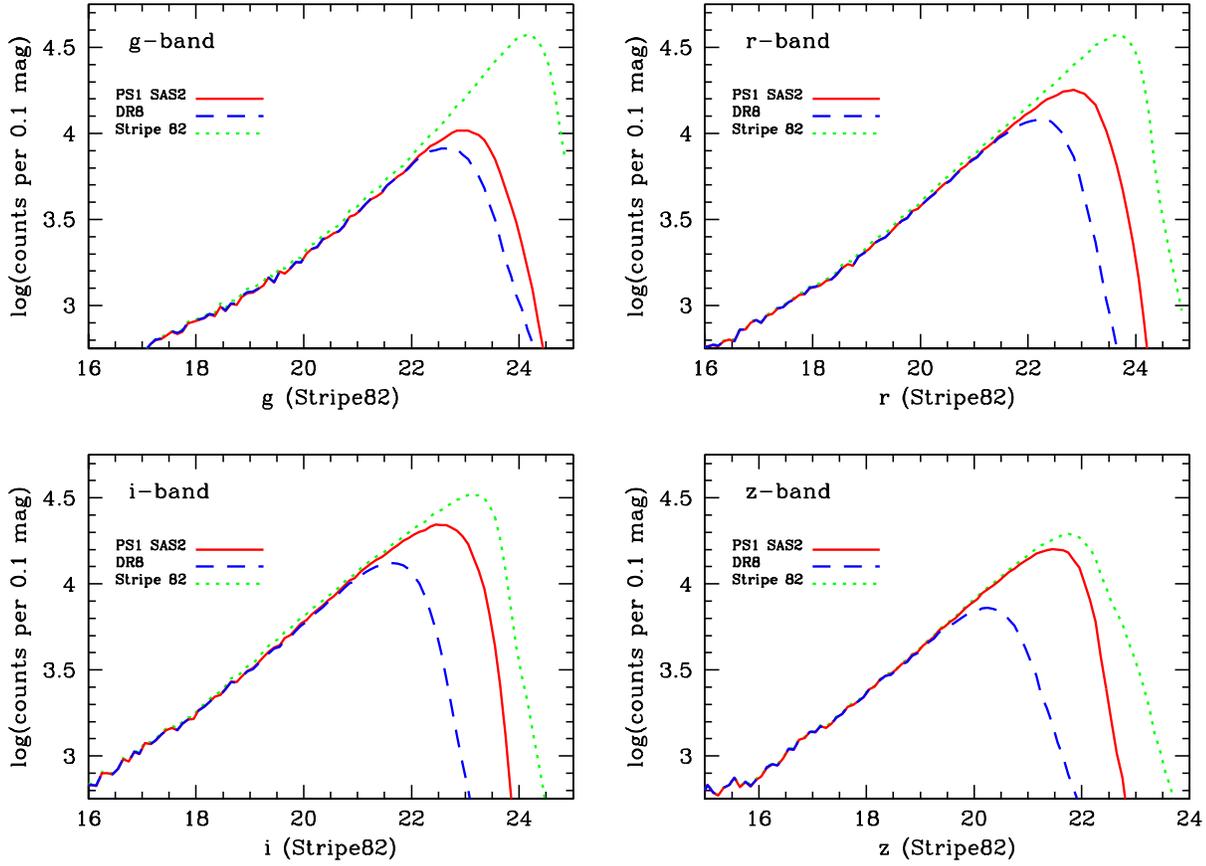}}
\caption{Differential object counts for the area in common to SAS2, DR8 and Stripe 82: green, short-dash SDSS Stripe 82; blue, long-dash SDSS DR8 with a match in Stripe 82; red, solid line PS1 stacked SAS2 with a match in Stripe 82. All three datasets are plotted as a function of Stripe 82 modelMag. Only 5-sigma detections ({\tt BINNED1} flag {\tt TRUE}) are included in the SDSS data.}
\label{fig:allcounts}
\end{center}
\end{figure*}

As an immediate indication of the depth of SAS2 and DR8 we match both to the 
deeper Stripe 82 data, assuming Stripe 82 to be correct (in fact there are clearly some `false' sources in the Stripe 82
catalogue, but these have no effect on the number of matched objects).
Again we use a circular match radius of $1.0$~arcsec. 
In the event of two (or more) objects in Stripe 82 being found inside this 
radius the brighter one is matched. The rms separation between the Stripe 82 and PS1 coordinates of all matched objects is $0.02\pm0.15$~arcsec\ in RA 
and $-0.04\pm0.15$~arcsec\ in Dec..

\begin{table*}
\setlength{\tabcolsep}{3pt}
\caption{Stacked SAS2 Kron magnitude offsets and scatter relative to SDSS Stripe 82 Pogson-corrected modelMags for $g$, $r$, $i$ and $z$-bands., 
for the whole area in common with Stripe 82. Colour corrections from \citet{photo} have been applied. The r-band data are shown in Fig. \ref{fig:kroncomp}. Offsets are 
in the sense PS1 - Stripe 82. The magnitude differences have been clipped at $\pm1.5$~mag, due to the presence of
some extreme outliers.}
\begin{center}
\begin{tabular}{lllllllll}
\hline
\hline
{\bf Stripe82 mag.} & \multicolumn{2}{c}{\bf Kron - $g$}& \multicolumn{2}{c}{\bf Kron - $r$}& \multicolumn{2}{c}{\bf Kron - $i$}& \multicolumn{2}{c}{\bf Kron - $z$}\\
(Pogson)&(galaxies)&(stars)&(galaxies)&(stars)&(galaxies)&(stars)&(galaxies)&(stars)\\
\hline
$16-17$  &&&&&&&$0.15\pm0.21$&$0.05\pm0.03$ \\
$17-18$   &$0.07\pm0.21$& $0.06\pm0.03$ &$0.05\pm0.20$& $0.02\pm0.03$&$0.10\pm0.19$& $0.05\pm0.04$&$0.17\pm0.18$& $0.06\pm0.05$\\
$18-19$   &$0.09\pm0.18$& $0.06\pm0.05$ &$0.05\pm0.17$& $0.02\pm0.05$&$0.10\pm0.17$& $0.05\pm0.05$&$0.18\pm0.21$& $0.08\pm0.06$\\
$18-20$   &$0.10\pm0.20$& $0.07\pm0.06$ &$0.06\pm0.17$& $0.03\pm0.06$&$0.12\pm0.18$& $0.07\pm0.07$&$0.22\pm0.24$& $0.10\pm0.09$\\
$20-21$   &$0.15\pm0.22$& $0.08\pm0.09$ &$0.09\pm0.21$& $0.05\pm0.10$&$0.16\pm0.24$& $0.09\pm0.10$&$0.23\pm0.28$& $0.12\pm0.15$\\
$21-22$   &$0.18\pm0.26$& $0.10\pm0.14$ &$0.13\pm0.27$& $0.07\pm0.15$&$0.18\pm0.29$& $0.10\pm0.17$&$0.18\pm0.36$& $0.12\pm0.31$\\
$22-23$   &$0.16\pm0.35$& $0.10\pm0.27$ &$0.11\pm0.34$& $0.07\pm0.27$&$0.13\pm0.36$& $0.09\pm0.33$&&\\
\hline
\end{tabular}
\end{center}
\label{tab:sas2scatter}
\end{table*}

Fig. \ref{fig:allcounts} shows the differential number counts of matched 
objects in $g$, $r$, $i$ and $z$ bands respectively, together with the Stripe 82 counts. In all
cases we plot against Stripe 82 modelMag. Plotting against psfMag would move all the points
$\sim0.3$~mag fainter (the relative depths would not change), which  simply reflects the fact that 
at the limiting depth of the SAS2 most objects are galaxies not stars, and are therefore not well measured by a 
PSF-fit magnitude. In all bands the SAS2 data are deeper than those from DR8, as indeed they should be,
as the increase in exposure time for the stacked SAS2 ($\sim500$~s {\it cf} $\sim50$~s) more than compensates for the difference in telescope 
aperture between PS1 and Apache Point ($1.8$~m {\it cf} $2.5$~m). The camera is also more red sensitive than that used by SDSS, 
resulting in larger gains in the redder bands. In fact, in the $z$-band, SAS2 is nearly as deep as 
Stripe 82. However, a note of caution should be employed for the $3\pi$ survey as a whole - as shown
in Section \ref{sec:sas2}, the redder bands in SAS2 have a much fainter sky than is typical for $3\pi$, and in the seeing is somewhere better than the $3\pi$ as a whole,  so the average limits may be some 0.3--0.4~mag brighter than implied here.

As there is no SDSS $y$-band, we cannot compare with Stripe 82 to determine the \yps \ depth. However, we
do have deeper data in the SAS2 area in the form of the PS1 Medium Deep Field 9 (MD09). The PS1 Medium Deep fields (of which
there are ten) are single pointings ($\sim7$~deg$^2$) which are visited nightly and have longer individual exposures
than the $3\pi$ (240~s for the $y$-band). The stacked $y$-band data on MD09 currently consists of over 100 of these exposures.
Fig. \ref{fig:ycounts} shows the counts matched to MD09 for both the SAS2 data, and, as a comparison, those from the UKIDSS LAS \citep{UKIDSS} DR9 release in this area.
All are plotted as a function of MD09 Kron magnitude (as measured using {\sc psphot}). The stacked SAS2 data are about $0.6$ magnitudes deeper than the UKIDSS LAS data, and show a count turnover at about \yps$=20.8$. It should be 
noted, however, that \yps, with $\lambda_{\rm eff}\sim0.96$~${\rm \mu m}$ is somewhat bluer that the UKIDSS $Y$-band, which stretches from 0.97--1.07~$\mu$m \citep{UKIDSSphoto}.

Apart from the $y$-band, we have presented our depth estimates as function of Stripe 82 modelMag. 
The question naturally arises how do these compare with the SAS2 Kron magnitudes? Fig. \ref{fig:kroncomp} shows the $r$-band magnitude comparison between 
the two systems (for clarity, we only plot a random subset of 25000 out of the $\sim$440000 objects in common). Colour terms between SDSS and PS1 systems
are taken from the linear relations in \citep{photo}, although for the $r$-band the correction is only of order $0.01(g-r)$. Strictly speaking these are only
appropriate for main sequence stars, but they should be representative for 
most galaxies.  
Brightward of $r\sim15.5$ saturation is an issue (probably in both datasets, but certainly in
Stripe 82 as these bright stars are classified as galaxies by SDSS). Apart from 
that the comparison appears quite reasonable. Table \ref{tab:sas2scatter} 
lists the  the magnitude offsets and scatter as a function of magnitude for
the $g$, $r$, $i$ and $z$ bands.
As we have discussed previously, Kron magnitudes are (by definition) not 
expected to be total, and the amount of light
lost should be larger for galaxies than for stars. Table \ref{tab:sas2scatter} seems to bear this 
out, if we make the assumption that modelMags are close to total, with all the offsets showing the Kron magnitude to be fainter, and the galaxy offsets generally $0.03-0.05$~mag larger than those for the stars in all the bins, apart
from the $z$ band where they are closer to $0.1$~mag. 
This is quite close to the theoretical expectations, given {\sc psphot} uses a 
Kron multiplier of 2.5 (see Section \ref{subsec:simul}). There is a slight trend for the offsets to become larger at fainter magnitudes for both stars and galaxies. As discussed in Section \ref{sec:sdsspixels}, we suspect this
is due to an underestimation of the Kron radius for faint objects. In an ideal noise-free world, where the summation for the 
Kron radii could be extended to infinite radius, this should not happen, but in the real world we consider a shift between the two systems of only $\sim0.05$~mag over a six magnitude range to be quite impressive. 

\begin{figure}
\begin{center}
\centerline{\includegraphics[width=3.5in]{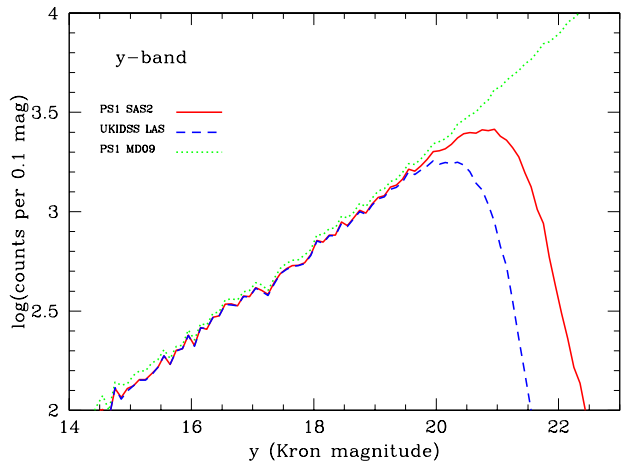}}
\caption{Differential object counts as a function of PS1 MD09 Kron magnitude for the area in common to MD09, SAS2 and UKIDSS LAS: green, dotted line MD09; blue, dashed line UKIDSS LAS with a match in MD09; red, solid line PS1 stacked SAS2 with a match in MD09.}
\label{fig:ycounts}
\end{center}
\end{figure}

We summarise the depth results from this section, and Section \ref{subsec:stack}, in Table \ref{tab:depth}.
For the count turnover magnitude we have used Stripe 82 modelMags, in order to aid the comparisons between the surveys,
except for the $y$-band, where we use PS1 Kron magnitudes from the Medium-Deep survey.
The 50 per cent point source completeness limits, being internal to PS1, are given in {\tt CAL\_PSF\_MAG} magnitudes. 
The offsets between the two are in line with the expectation from Figs \ref{fig:limit50} and \ref{fig:numdepth}.

\begin{table}
\setlength{\tabcolsep}{3pt}

\caption{Measured depths of the various surveys on the SAS2 area, as judged by the peak of the differential number-magnitude counts.
The $g, r, i, z$ PS1 and DR8 data only include objects with a match in Stripe 82. Both the SDSS catalogues are restricted
to objects of $5\sigma$ significance or more ({\tt BINNED1} flag {\tt TRUE}) for the particular band. Also shown are the 50 per cent detection limits for point sources on the PS1 data.
The PS1 results come from the stacked data. All magnitudes are SDSS Stripe 82 modelMags unless otherwise noted.}

\begin{center}
\begin{tabular}{cclccl}
\hline
\hline
{\bf Band} & {\bf PS1} & {\bf PS1} & {\bf DR8} & {\bf Stripe 82}&\bf{UKIDSS LAS}\\
& {\bf (50\%)} & \multicolumn{4}{c}{\bf (count turnover magnitude)}\\

\hline
$g$ &$23.4^*$ & 23.0 & 22.8&24.2&\\
$r$ &$23.4^*$ & 22.8 & 22.2 & 23.6&\\
$i$ &$23.2^*$ & 22.5 & 21.6 & 23.1&\\
$z$ &$22.4^*$ & 21.7 & 20.3 & 21.8&\\
$y$ &$21.3^*$ & $20.8^{**}$ &&&$20.2^{**}$\\
\hline
\multicolumn{6}{l}{$^*$ PS1 SAS2 PSF magnitude }\\
\multicolumn{6}{l}{$^{**}$ PS1 Kron magnitude from Medium Deep Field 9 }\\
\end{tabular}
\end{center}
\label{tab:depth}
\end{table}

\begin{figure}
\begin{center}
\centerline{\includegraphics[width=3.75in]{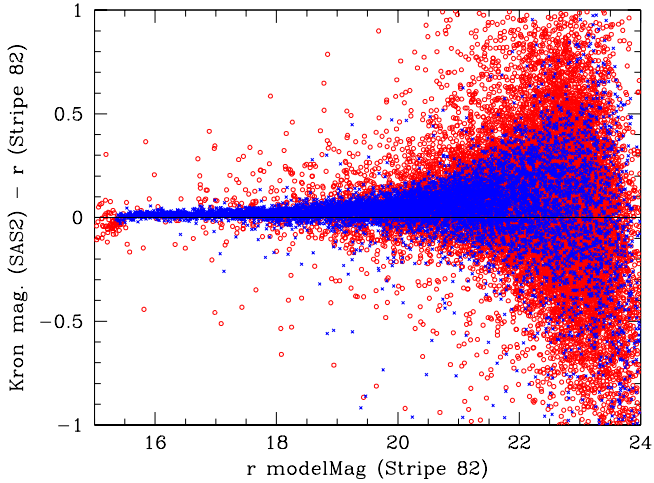}}
\caption{The difference between stacked SAS2 Kron magnitude and SDSS Stripe 82 modelMag as a function of the SDSS magnitude, for a random sample of all objects in common.  Objects 
classified as stars by SDSS (type = 6) are blue crosses, galaxies (type = 3) are red circles. Table \ref{tab:sas2scatter} presents the results for the whole dataset in numerical form.}
\label{fig:kroncomp}
\end{center}
\end{figure}

\section{False detections}
\label{sec:junk}

The PS1 camera is essentially a prototype, designed for fast 
readout and  charge shuffling (although the latter has not been implemented 
for the PS1 surveys),
and does suffer from a variety of defects, many of which show up as false 
detections. This is not helped by the large number of detector edges which 
come from having nearly 4000 individual CCD cells. Also, with so many 
detectors, the decision was taken to use slightly imperfect chips, 
resulting in a very large saving of both cost and manufacturing time.

One particular problem has been the issue of variable dark/bias signal,
which can alter on the timescale of single exposures, and on
certain CCDs on a spatial scale right down to single rows. This can make 
accurate subtraction a challenge.  We believe that the two spikes seen in the 
PS1 power spectra (Fig. \ref{fig:power}) are related to this issue.
There is also cross-talk between certain OTAs, persistence trails
left by bright stars, and ghost images due to reflections. 
Efforts are ongoing to alleviate these problems.
As far as the image detection software is concerned, looking back at 
Fig. \ref{fig:sdsstile_count} it is clear that, when run 
on the same pixels, {\sc psphot} is no worse than SDSS at picking 
up false objects.

\begin{figure}
\begin{center}
\centerline{\includegraphics[width=3.5in]{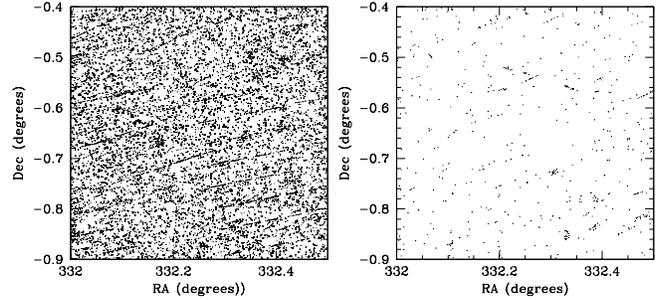}}
\caption{The location of $g$-band false detections on the sky for a typical $0.5\degr\times0.5\degr$ area of SAS2. Left: the combination of all objects detected in all the individual exposures; right: the same area but now only for the stacked data.}
\label{fig:false_on_sky}
\end{center}
\end{figure}

Of course, the outlier clipping applied during the stacking process would be 
expected to remove, or at least dilute the 
effect of, many of the defects. This is demonstrated in Fig. \ref{fig:false_on_sky} which shows the 
locations of PS1 objects with no match in Stripe 82 on a typical area of SAS2,
both for all the individual exposures, and for the stacked data. The linear, 
diagonal feature in the individual exposures are due to defects at the edges
of individual CCD cells (which often correlate between adjacent cells) which have not
been masked. Most of these features disappear in the stack, as the defects 
in the individual exposures to not line up on the sky (due to dithering and
rotation).

We investigate here how these problems affect the number of 
detections on the SAS2 stacks. To do this we return to the sample matched to Stripe 82 
described in 
Section \ref{sec:surveys}, but we now also consider the objects in SAS2 
which have no match in Stripe 82. We exclude from this sub-sample all objects 
with {\tt PSF\_QF\_PERFECT} $<0.85$ (which removes objects with more than 15 per cent of masked pixels, weighted by the PSF,  
whose positions may be inaccurate) and with any of the following {\sc psphot} analysis FLAGS set:   
{\tt FITFAIL, SATSTAR, BADPSF, DEFECT, SATURATED, CR\_LIMIT, MOMENTS\_FAILURE, SKY\_FAILURE, SKYVAR\_FAILURE, SIZE\_SKIPPED}, which correspond to a hex flag value 
of 0x1003bc88. These are mainly objects for which the 
software has failed in some way, and so measurements are unreliable \citep[see Table 2 of][]{magnier12}. This 
reduces the number of false detections (in all bands) by about 20-25 per cent.
Remember we have already removed from the sample areas around very 
bright stars where there are holes in the Stripe 82 catalogue. It is likely that there are a 
significant number of false PS1 detections in these areas (this is not an issue unique
to PS1, of course, and is presumably why there are holes in the Stripe82 catalogue in the first place).
We will return to the issue of false detections around bright stars in Paper II, where we design
 a mask for the survey based on the positions and magnitudes of known stars. Here, we are more concerned with those defects which
are peculiar to PS1 and the way the camera is constructed.

Figs \ref{fig:gfalsepsf} and \ref{fig:gfalsekron} show
the differential number counts of objects in the \rps-band, for all objects, 
and for those objects with and without matches in the SDSS Stripe 82 catalogue.
As might be expected, the PSF unmatched counts rise sharply towards the 
limiting magnitude of the data, as noise spikes (and other background artefacts) 
start to be detected as real objects.
The Kron false counts, however, tend to be more spread out, and have a lower peak 
(note that the integrated number of false detections is very similar - only about 
5 per cent are lost due to a failure to determine a Kron magnitude).
Some of this is just due to errors, but we believe a significant number 
of the defects, detected at low significance 
with the PSF fits, are extended, and so grow significantly brighter 
when measured with a Kron technique. 

\begin{figure}
\begin{center}
\centerline{\includegraphics[width=3.75in]{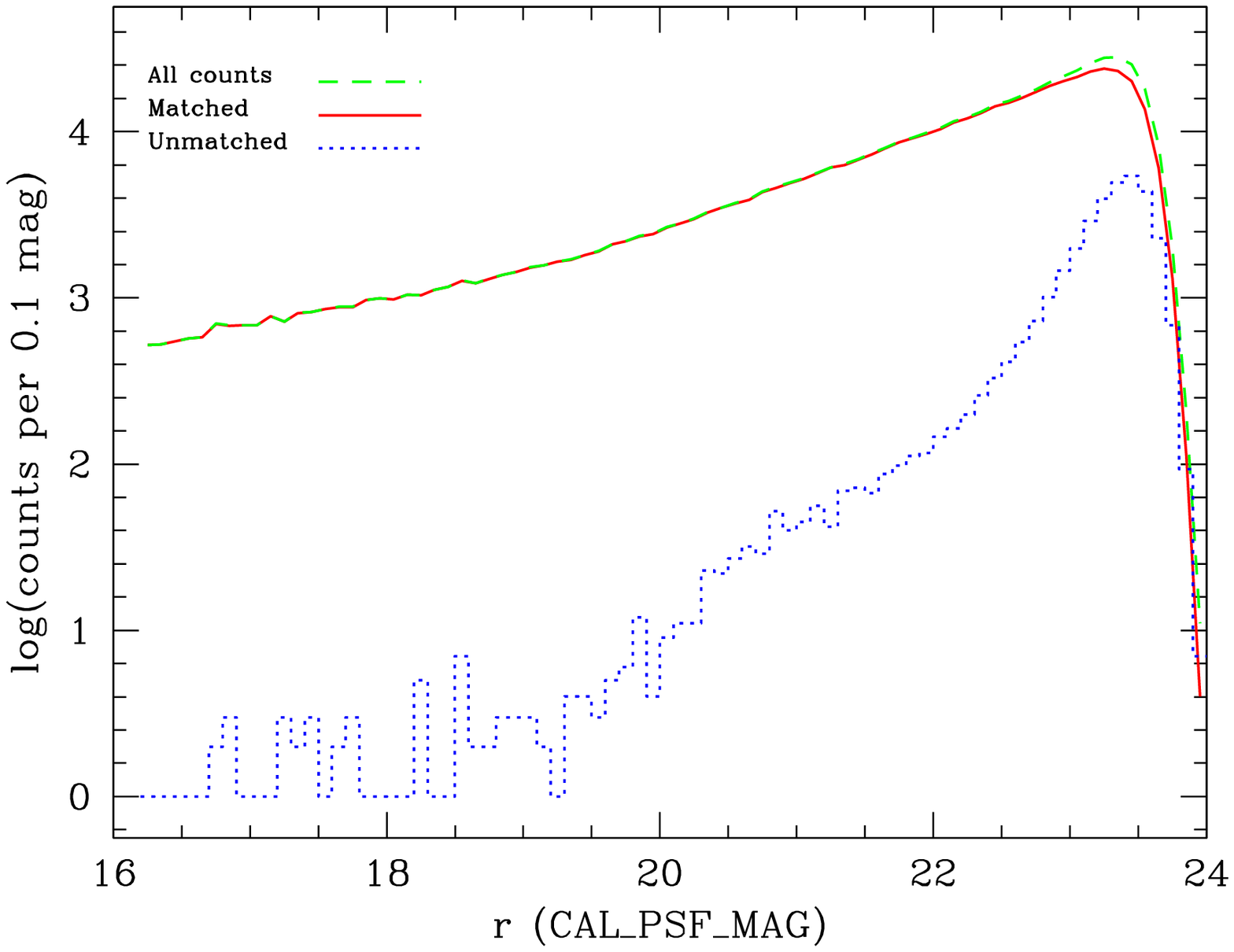}}
\caption{Differential \rps-band counts, as a function of PS1 {\tt CAL\_PSF\_MAG} magnitude, for all objects (dashed green line), those matched Stripe 82 (solid red line) and those not matched to Stripe 82 (dotted blue line). Only 5-sigma detections ({\tt BINNED1} flag {\tt TRUE}) are included in the SDSS data. Objects with any of the flags listed in the text
set are excluded from the PS1 data.}
\label{fig:gfalsepsf}
\end{center}
\end{figure}

\begin{figure}
\begin{center}
\centerline{\includegraphics[width=3.75in]{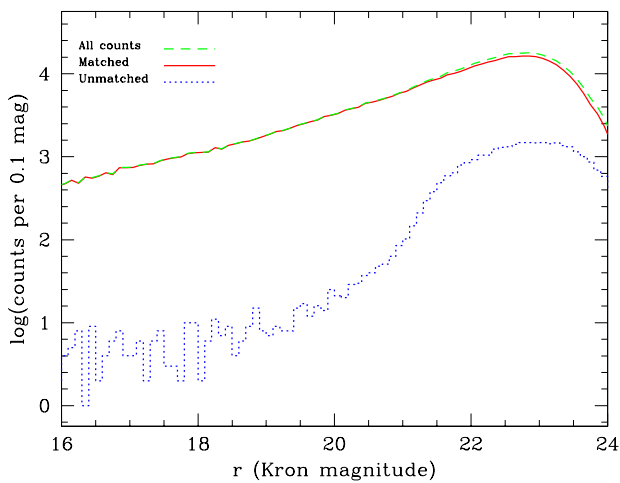}}
\caption{As Fig. \ref{fig:gfalsepsf}, but now as a function of PS1 Kron magnitude.}
\label{fig:gfalsekron}
\end{center}
\end{figure}

\begin{figure*}
\begin{center}
\centerline{\includegraphics[width=6.75in]{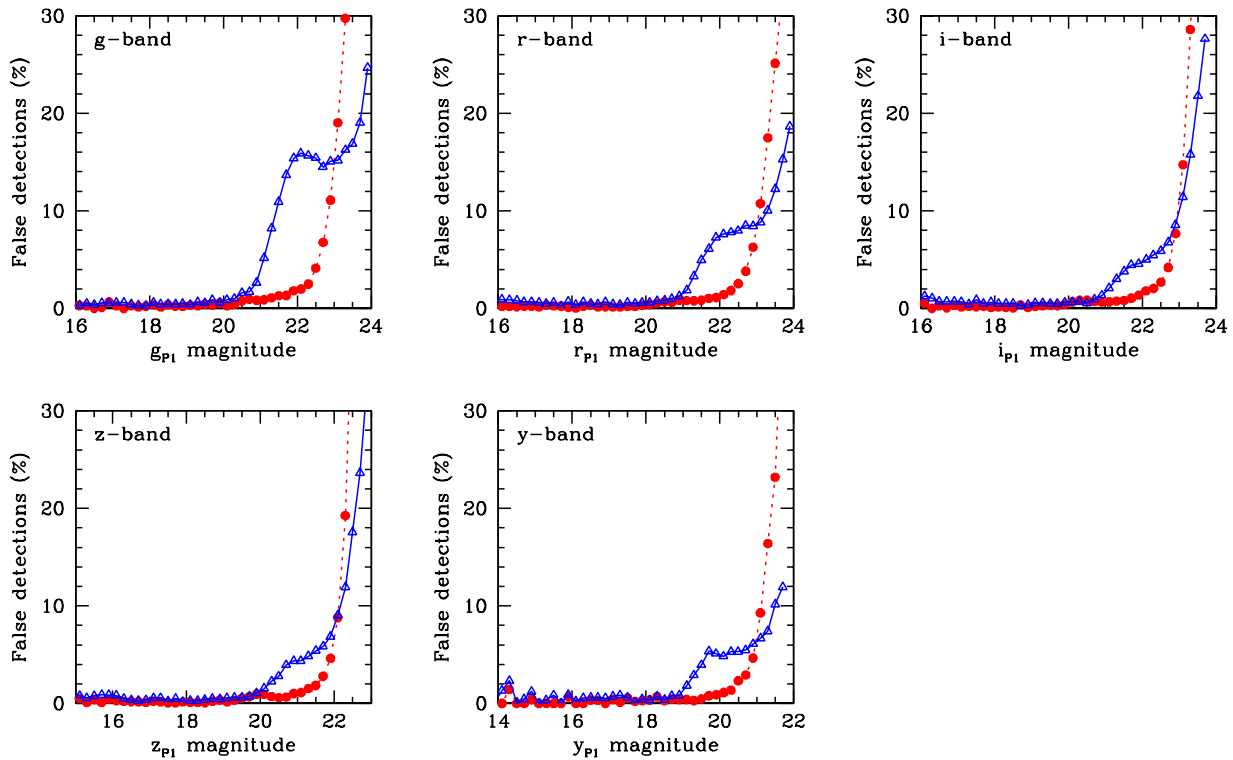}}
\caption{The percentage of detections from the stacked SAS2 data with no match to Stripe 82 (\gps, \rps, \ips, \zps) or PS1 Medium Deep Field 9 (\yps) as a function of both PSF magnitude (red circles) and Kron magnitude (blue triangles). Objects with any of the PS1 flags set which are listed in the text are excluded.}
\label{fig:falsepercent}
\end{center}
\end{figure*}

The difference between the two magnitude systems is highlighted in Fig. \ref{fig:falsepercent}, which shows the percentage of false detections as a
function of both PSF and Kron magnitudes, for \gps, \rps, \ips, \zps and \yps 
bands. For \yps, where there is no corresponding SDSS band, we have matched 
to the deeper PS1 data on Medium Deep Field 9.
The \gps-band is clearly the worst - at the PSF magnitude corresponding 
to the
50 per cent completeness limit from Table \ref{tab:depth} around 39 per cent 
of detections are false.  For the Kron magnitude turnover the situation 
is not as bad, with only 15 per cent of objects being false, and in the 
\rps band the corresponding figures have dropped to 25 per cent and 
8 per cent. This 
may reflect the fact that the \gps \ exposures are more dominated by read noise
than the other bands, due to the lower \gps band sky. It may also be in
some part due to ghosts caused by bright stars reflecting off the surface 
of the detector and back off the coating on the underside of one of the 
correctors. These ghosts are known to strongly favour shorter wavelengths 
(they are virtually undetectable in \ips, \zps \ or \yps). In principle, 
their locations can be predicted, so it should be possible to mask out 
most of the affected areas. Some of this is already done for the brightest 
stars.

\section{Discussion and Conclusions}

PS1 not only uses a unique camera but relies on a purpose-built software 
pipeline to reduce the data. We have shown, by creating fake exposures, and by
adding fake objects to real exposures, that the pipeline works as expected,
and that the warping and stacking processes are well behaved. The depth of 
the data also scales correctly with exposure time. As a further check,
we have run the pipeline on SDSS fields and recovered very similar magnitudes
and numbers of objects to those in the SDSS catalogues.

By matching both PS1 and SDSS DR8 datasets to SDSS Stripe82, we have determined that the SAS2 PS1 data are deeper than SDSS DR8 by $\sim0.2$, $\sim0.6$, $\sim0.9$ and $\sim1.4$~mag  
in $g$, $r$, $i$ and $z$ respectively.  The $z$ depth is
within $0.1$~mag of that of SDSS Stripe 82. As we have no external deeper $y$-band data
on this field, we have had to perform an internal comparison with the PS1 Medium Deep data. We
find that \yps is $\sim0.6$~mag deeper than the UKIDSS LAS.

PSF and Kron magnitudes are being measured reliably, and agree well with SDSS,
apart from a slight magnitude dependent scale error in the Kron magnitudes.
This results in PS1 magnitudes becoming systematically too faint 
with decreasing 
flux, by $\sim0.05$~mag over a $6$ magnitude range. We suspect this is due to
a slight underestimation of the Kron radius at faint magnitudes, although
why our reduction of the SDSS fields shows a much larger effect is a puzzle.
The scatter between PS1 and Stripe 82 ranges from $\pm0.02$ for the brightest objects in common, to about $\pm0.3$~mag at the limit of the PS1 data.

False positives are still something of an issue for PS1, but, using the 
default $5\sigma$ detection threshold, are still under 15 per cent in all bands at 
the limiting Kron magnitude 
of the survey. The reduction of the SDSS fields showed a very similar number of false
detections to SDSS DR8 itself, so the problems lie with the data itself 
not the 
software. The fact that the false positive rates are relatively higher in 
the \gps\ band might be indicative that a proportion of these false detections 
are wavelength-dependent ghost reflections from bright stars. The false detection rate could be further reduced by insisting that objects exist in at least two bands, although this would be at the expense of
limiting magnitude, and may preclude the discovery of faint, `drop-out' 
galaxies in the redder bands. There may also be some benefit to performing
forced photometry on the {\it individual} exposures at the locations of objects
detected on the stacks - presumably the false detections would show inconsistent results between the exposures.

It has to be borne in mind that the SAS2 probably represents some of the best 
conditions that will be found in the $3\pi$ survey. It was taken under mostly 
dark sky conditions, even in the redder bands, which are normally taken during grey or bright time, and the seeing was $0.1-0.2$~arcsec better than the median of the existing $3\pi$ exposures. 
As a result, the average depth of the final stacked $3\pi$ survey cannot expected to be as
good as SAS2.

In paper II we will present a simple star/galaxy separation method, calibrated 
using our synthetic images, and attempt to quantify the effect of the 
spatially varying depth across the SAS2 on the counts and angular clustering 
of galaxies.

For this paper we have run our own instance of the PS1 software on the
pixel data, based on a build of {\sc psphot} from September 2012 (software version number 34471).
The data which will be released to the user community will be in the form 
of database access to catalogues generated by the pipeline in Hawaii. 
To ensure consistency, we have run extensive comparisons between our 
results and those currently available for SAS2 from 
Hawaii and find virtually identical results, so we are confident that the 
conclusions presented here will also apply to the initial released 
catalogues (the first release of the $3\pi$ survey is to be based on 
virtually the same pipeline code as SAS2). 

\section*{Acknowledgements}

The Pan-STARRS1 Surveys (PS1) have been made possible through
contributions of the Institute for Astronomy, the University of
Hawaii, the Pan-STARRS Project Office, the Max-Planck Society and its
participating institutes, the Max Planck Institute for Astronomy,
Heidelberg and the Max Planck Institute for Extraterrestrial Physics,
Garching, The Johns Hopkins University, Durham University, the
University of Edinburgh, Queen's University Belfast, the
Harvard-Smithsonian Center for Astrophysics, the Las Cumbres
Observatory Global Telescope Network Incorporated, the National
Central University of Taiwan, the Space Telescope Science Institute,
the National Aeronautics and Space Administration under Grant
No. NNX08AR22G issued through the Planetary Science Division of the
NASA Science Mission Directorate, the National Science Foundation
under Grant No. AST-1238877, and the University of Maryland.
Durham University's membership of PS1 was made possible through the
generous support of the Ogden Trust.

Funding for the SDSS and SDSS-II has been provided by the
Alfred P. Sloan Foundation, the Participating Institutions, 
the National Science Foundation, the U.S. Department of
Energy, the National Aeronautics and Space Administration,
the Japanese Monbukagakusho, the Max Planck Society,
and the Higher Education Funding Council for England.
The SDSS Web Site is http://www.sdss.org/.
This work is based in part on data obtained as part of the UKIRT 
Infrared Deep Sky Survey.

PN acknowledges the support of the Royal Society
through the award of a University Research Fellowship, and the 
European Research Council through receipt of a Starting 
Grant (DEGAS-259586). PWD acknowledges support from the ERC Starting Grant
(DEGAS-259586).

\label{lastpage}


\begin{thebibliography}{99}
\bibitem[\protect\citeauthoryear{Abazajian et al.}{2009}]{sdss7} Abazajian K.~N. et al. (The SDSS Collaboration),  2009, ApJS, 812, 543

\bibitem[\protect\citeauthoryear{Aihara et al.}{2011}]{sdss8} Aihara H., Allende Prieto C., An D., et al., 2011, ApJS, 193, 29

\bibitem[\protect\citeauthoryear{Annis et al.}{2011}]{s82} Annis L., Soares-Santos M., Strauss M.~A. et al., 2011, arXiv:1111.6619

\bibitem[\protect\citeauthoryear{Bertin \& Arnouts}{1996}]{sex} Bertin E., Arnounts S., 1996, Astr. Astroph. Suppl. 117, 393

\bibitem[\protect\citeauthoryear{Farrow et al.}{2013}]{paper2} Farrow D.~J. et al., 2013,  MNRAS Accepted (Paper II)

\bibitem[\protect\citeauthoryear{Hewett et al.}{2006}]{ukidss} Hewett P.~C., Warren S.~J., Leggett S.~K., Hodgkin S.~T., 2006, MNRAS 367, 1521

\bibitem[\protect\citeauthoryear{Hodapp et al.}{2004}]{hodapp} Hodapp K.~W., Siegmund W.~A., Kaiser N., Chambers K., Laux U., Morgan J., Mannery E., 2004, Proc. SPIE 5489, 667

\bibitem[\protect\citeauthoryear{Hodgkin et al.}{2009}]{UKIDSSphoto} Hodgkin S.~T., Irwin M.~J., Hewett P.~C., Warren S.~J., 2009, MNRAS 394, 675

\bibitem[\protect\citeauthoryear{Kaiser et al.}{2010}]{PS1_system} Kaiser N. et al., 2010, Proc. SPIE, 7733

\bibitem[\protect\citeauthoryear{Kron}{1980}]{kron} Kron R.~G., 1980, ApJ Suppl. 43, 305

\bibitem[\protect\citeauthoryear{Lawrence et al.}{2007}]{UKIDSS} Lawrence A. et al., 2007, MNRAS, 379, 1599

\bibitem[\protect\citeauthoryear{Magnier et al.}{2006}]{PS1_IPP} Magnier E., Kaiser N., Chambers K., 2006, Proceedings of The Advanced Maui Optical and Space Surveillance Technologies Conference, Ed.: S. Ryan, The Maui Economic Development Board, p. 455-461

\bibitem[\protect\citeauthoryear{Magnier et al.}{2008}]{PS1_astrometry} Magnier E.~A., Liu M., Monet D.~G., Chambers K.~C., 2008, IAU Symposium,  248, 553

\bibitem[\protect\citeauthoryear{Magnier et al.}{2013}]{magnier12} Magnier E.~A. et al., 2013, ApJ, in press

\bibitem[\protect\citeauthoryear{Merson et al.}{2013}]{merson} Merson A.~I. et al., 2013, MNRAS, 429, 556

\bibitem[\protect\citeauthoryear{Onaka et al.}{2008}]{onaka} Onaka P., Tonry J.~L., Isani S., Lee A., Uyeshiro R., Rae C., Robertson L., Ching G., 2008, Proc. SPIE 7014, 12O

\bibitem[\protect\citeauthoryear{Schlafly et al.}{2012}]{ubercal} Schlafly E.~F. et al., 2012, ApJ 756, 158

\bibitem[\protect\citeauthoryear{Schlegel et al.}{1998}]{s98} Schlegel D.~J., Finkbeiner D.~P., Davis N., 1998, ApJ, 500, 525

\bibitem[\protect\citeauthoryear{Shen et al.}{2003}]{shen} Shen S., Mo H.~J., White S.~D.~M., Blanton M.~R., Kauffmann G., Voges W., Brinkmann J., Csabai I., 2003, MNRAS, 343, 978

\bibitem[\protect\citeauthoryear{Tonry et al.}{2008}]{camera} Tonry J., Burke B., Isani S., Onaka P., Cooper M., 2008, Proc. SPIE, 7021

\bibitem[\protect\citeauthoryear{Tonry \& Onaka}{2009}]{PS1_GPCA} Tonry J., Onaka P., 2009, Proceedings of the Advanced Maui Optical and Space Surveillance Technologies Conference, Ed.: S. Ryan, p.E40.


\bibitem[\protect\citeauthoryear{Tonry et al.}{2012}]{photo} Tonry J.~L. et al.,2012, ApJ 750, 99

\bibitem[\protect\citeauthoryear{York et al.}{2000}]{sdss} York D. G., et al., 2000, AJ 120, 1579

\end{thebibliography}
\end{document}